%% file: main.tex
\pdfoutput=1
\documentclass[journal]{IEEEtran}

\usepackage{graphicx}

\usepackage{caption}
\usepackage{threeparttable}
\usepackage{subfigure}
\usepackage{tabularx}
\usepackage{booktabs}
\usepackage{lipsum}
\usepackage{multirow}
\usepackage[usenames,dvipsnames]{color}
\usepackage{colortbl}
\definecolor{mygray}{gray}{.9}

\usepackage{amsmath,bm}

\usepackage{algorithm,algpseudocode}
\usepackage{amsmath}
\usepackage{amsfonts}
\usepackage{amssymb}

\usepackage{cite}

\usepackage{nomencl}
\makenomenclature

\begin{document}
\captionsetup[figure]{labelfont={default},labelformat={default},labelsep=period,name={Fig.}}
\captionsetup{justification   = raggedright, singlelinecheck = false}
\title{WAMS-Based Model-Free Wide-Area Damping Control by Voltage Source Converters}
%
%
%

\author{Jinpeng Guo,~\IEEEmembership{Student Member,~IEEE,}
Ilias Zenelis,~\IEEEmembership{Student Member,~IEEE,}
        Xiaozhe Wang,~\IEEEmembership{Member,~IEEE,}
        Boon-Teck Ooi,~\IEEEmembership{Life~Fellow,~IEEE}
\thanks{This work was supported by the Natural Sciences and Engineering Research Council (NSERC) under Discovery Grants RGPIN-2016-69152 and RGPIN-2016-04570, and the Fonds de Recherche du Quebec-Nature et technologies under Grant FRQ-NT PR-253686. }
\thanks{ J. Guo, I. Zenelis, X. Wang and B. Ooi are with the Department of Electrical and Computer Engineering, McGill University, 3480 University Street, Montreal, Canada. (email:  jinpeng.guo@mail.mcgill.ca; ilias.zenelis@mail.mcgill.ca; xiaozhe.wang2@mcgill.ca; boon-teck.ooi@mcgill.ca)}
}

%
%



\maketitle

\begin{abstract}
In this paper, a novel model-free wide-area damping control (WADC) method is proposed, which can achieve full decoupling of modes and damp multiple critical inter-area oscillations simultaneously using grid-connected voltage source converters (VSCs). The proposed method is purely measurement-based and requires no knowledge of the network topology and the dynamic model parameters. Hence, the designed controller using VSCs can update the control signals online as the system operating condition varies. Numerical studies in the modified IEEE 68-bus system with grid-connected VSCs show that the proposed method can estimate the system dynamic model accurately and can damp inter-area oscillations effectively under different working conditions and network topologies. 

\end{abstract}

\begin{IEEEkeywords}
phasor measurement units, voltage source converters, wide-area damping control, wide-area measurement system
\end{IEEEkeywords}

%
\IEEEpeerreviewmaketitle

\mbox{}

\nomenclature[1]{$M$}{Generators' inertia coefficient matrix}
\nomenclature[2]{$D$}{Generators' damping coefficient matrix}
\nomenclature[3]{$\bm{\delta}$}{The vector of generators' rotor angles}
\nomenclature[4]{$\bm{\omega}$}{The vector of generators' rotor speeds}
\nomenclature[5]{$\bm{P_E}$}{The vector of active power injections from the buses where generators are connected}
\nomenclature[6]{$E$}{A diagonal matrix whose diagonal components are generators' voltage magnitude behind transient reactance} 
\nomenclature[7]{$G$}{A diagonal matrix whose diagonal entries are the diagonal elements of the reduced conductance matrix\color{black}} 
\nomenclature[8]{$\Sigma$}{The load fluctuation intensity matrix} 
\nomenclature[9]{$\bm{\xi}$}{A standard Gaussian random vector}
\nomenclature[a0]{$K_1$}{The damping coefficient matrix for VSCs' active power control}
\nomenclature[a1]{$K_2$}{The damping coefficient matrix for VSCs' reactive power control}
\nomenclature[a2]{$\bm{P_v}$}{The active power injections from the buses where VSCs are connected}
\nomenclature[a3]{$\bm{Q_v}$}{The reactive power injections from the buses where VSCs are connected}
\nomenclature[a4]{$\bm{\theta}$}{The vector of voltage angles of the buses where VSCs are connected} 
\nomenclature[a5]{$\bm{V}$}{The vector of voltage magnitudes of the buses where VSCs are connected} 

\nomenclature[a6]{$A$}{System state matrix in the state-space model}
\nomenclature[a7]{$B$}{System input matrix in the state-space model}
\nomenclature[a8]{$S$}{Noise input matrix in the state-space model}
\nomenclature[a9]{$A_c$}{Closed-loop system state matrix}
\nomenclature[b1]{$A_o$}{Open-loop system state matrix}
\nomenclature[b2]{$\bm{x}$}{System state variables}
\nomenclature[b3]{$\bm{z}$}{System modal variables}
\nomenclature[b4]{$R(\tau)$}{Stationary $\tau$-lag correlation matrix} 
\nomenclature[b5]{$C$}{Stationary covariance matrix}
\nomenclature[b6]{$J_C$}{The quadratic cost function in MLQR control}
\nomenclature[b7]{$W_Q$}{The weighting matrix for mode cost in MLQR control} 
\nomenclature[b8]{$W_R$}{The weighting matrix for input cost in MLQR control} 
\nomenclature[b9]{$\Gamma$}{The feedback gain matrix derived from MLQR control} 

\printnomenclature

\section{Introduction}
%
%
%
%

\IEEEPARstart{P}{ower} system oscillations have been traditionally damped by employing the power system stabilizer (PSS) \cite{PSS1} and Flexible AC Transmission System (FACTS) \color{black} with Power Oscillation Damping (POD) controller \color{black} \cite{STATCOM1}. However, PSS might be ineffective in damping multiple inter-area oscillations \cite{ref:LQG}, while {\color{black} the damping performance of both PSS and FACTS-POD controller may vary with their locations \cite{STATCOM1}}. More recently, {\color{black} growing renewable energy sources (RES) are integrated into the power grids. The inverter-based RES generators may interact with the damping torque, resulting in either improved or deteriorated system stability depending on different working conditions \cite{ref:quintero2014impact,ref:zhou2017damping}. Meanwhile, considering the flexible controls of inverter-based generations, many efforts have been made to develop effective POD controllers by exploiting the inverter-based generations \cite{ref:tang2015sliding, ref:offshore-wind, ref:multi-terminal-wind}.} Particularly, voltage source converters (VSCs), serving as the interface between RES and the AC power system, are regarded as new and promising candidates to improve AC system dynamic performance.  Compared to the {\color{black} conventional } FACTS devices that can control only reactive power, VSCs connected with RES can control both active and reactive power independently, which provides more flexibility in control design \cite{ref:yazdani2010voltage}. {\color{black} Note that several FACTS devices equipped with energy storage systems can also control active power, but they share similar features in topology and functions with VSC integrating RES \cite{ref:beza2014adaptive}. Therefore, we mainly focus on VSCs for POD in this paper, whereas it is believed that other FACTS devices similar to VSCs can be exploited in the same fashion. } \color{black}

{\color{black}Many control methods} have been proposed to enhance the system \color{black}POD \color{black} performance using VSCs such as {modal linear quadratic} Gaussian (LQG) {control} \cite{ref:LQG}, the traditional PSS-based { control} \cite{ref:trinh2014methods}, {mixed} $H_2/H_{\infty}$ output feedback control \cite{ref:mixed-H},  {Lyapunov control theory} \cite{ref:multi-terminal-wind} and sliding mode robust control \cite{ref:tang2015sliding}. \color{black}Although the robust control methods (e.g., mixed $H_2/H_{\infty}$ feedback control) provide the designed controllers with some robustness to the uncertainties of parameters and noise, \color{black} all the aforementioned methods require accurate knowledge of network topology and system parameters when the controllers are designed offline. However, the network may suffer from undetected topology changes 
and the system parameters may vary in different operation conditions. Therefore, effectiveness of the controllers designed offline may be deteriorated. { Such problems are aggravated after the integration of RES.} To address these issues, wide-area measurement system (WAMS)-based wide-area damping control (WADC) methods have been proposed in \cite{ref30}, \cite{ref31}. However, the method proposed in \cite{ref30} is not purely model-free, requiring the damping coefficients. Also, the method may not be directly implemented by VSCs. {\color{black}Besides, the method proposed in \cite{ref31} may not optimize the damping performance of multiple modes simultaneously.} {Several online WADC  methods have been discussed in \cite{ref:zhang2016review}, which nevertheless also points out the gap between the online identification and control, as the reduced-order model obtained from the system identification methods may not be the physical model { and cannot be related to the real system state variables}, making the control design challenging.}

In this paper, we propose a novel model-free WADC method using VSCs, {\color{black}which can achieve full decoupling of modes such that the damping performance of multiple inter-area modes can be optimized} simultaneously as the system \color{black} steady-state \color{black} operating condition varies. Specifically, we first integrate the model of VSCs into the dynamic AC power system model in the form of state-space representation. Next, a perturbation approach is designed to estimate the state matrix $A$ and the input matrix $B$ only from PMU data by exploiting the regression theorem of multivariate Ornstein-Uhlenbeck process \cite{regression-theorem}. It should be noted that, unlike other measurement-based mode identification methods (e.g., system identification methods), the estimated $A$ and $B$ {\color{black} correspond to the linearization of the \textit{physical model}, which possesses clear physical interpretations} and can be directly utilized in control design. Therefore, the  {\color{black} modal linear quadratic regulator (MLQR)}-based control method is applied to the estimated $A$ and $B$ to design a wide-area damping controller using VSCs. 

To our best knowledge, the proposed method seems to be \textit{the first} model-free WADC strategy that can achieve full decoupling of modes and can target all critical modes simultaneously in various operation conditions and network topologies. The contributions of the paper are summarized below:
\begin{itemize}
\item An entirely model-free method is designed to estimate the system state matrix $A$ and the input matrix $B$ in the state-space representation of the \textit{true physical} model of a power system with VSCs. Compared to \cite{ref:sheng2019online}, additional model formulation and mathematical manipulation are conducted for the integration of VSCs. A perturbation approach is also designed to separate $A$ and $B$ from the closed-loop system state matrix. 
\item {Based on the estimated matrices, a MLQR-based WADC method, requiring no knowledge of network model, is proposed to update the control signals of VSCs so that multiple critical inter-area modes can be damped simultaneously without affecting others as the system operating condition varies, as opposed to the model-based WADC methods \cite{ref:trinh2014methods, ref:mixed-H, ref:LQG,ref:multi-terminal-wind,ref:tang2015sliding}.}
\end{itemize}

The rest of the paper is organized as follows. Section II presents the state-space representation of the AC system with grid-connected VSCs. Section III describes the WAMS-based estimation strategy for the system state matrix $A$ and the input matrix $B$ in the state-space representation. Section IV shows the MLQR-based WADC based on the estimated matrices using VSCs. Section V validates the proposed estimation and control strategy through comprehensive numerical studies. Section VI summarizes the conclusions. 

\section{System modeling}

\subsection{The Stochastic Model of AC power system}
In  this  paper,  we  consider  the small-signal  electromechanical stability around the steady state, in which the rotor angle dynamics dominate. Therefore, the classical generator dynamic model is considered:
\color{black}
\begin{equation}
\begin{aligned}
\mathop {\bm{\dot \delta }}   &= {\omega _0}\left( {{\bm{\omega }} - \bm{1}} \right)\\
{{M}}\mathop {\bm{\dot \omega }}   &= \bm{P_M} - \bm{P_E} - {{D}}\left( {{\bm{\omega }} - \bm{1}} \right)
\end{aligned} \label{eq1}
\end{equation}
\color{black}
with
\begin{equation}
{P_{ei}} = \sum\limits_{j = 1}^{{N_g}} {{E_i}{E_j}\left( {{G_{ij}}\cos \left( {{\delta _i} - {\delta _j}} \right) + {B_{ij}}\sin \left( {{\delta _i} - {\delta _j}} \right)} \right)} \nonumber
\end{equation}
where ${\bm{{\delta }}} = {\left[ {{\delta _1},{\delta _2},...,{\delta _{{N_g}}}} \right]^T}$ is the vector of generator rotor angles, ${\bm{\omega }} = {\left[ {{\omega _1},{\omega _2},...,{\omega _{{N_g}}}} \right]^T}$ is the vector of generator rotor speeds and ${{{\omega }}_{{0}}}$ is the base value, ${{M}} = diag\left( {\left[ {{M_1},{M_2},...,{M_{{N_g}}}} \right]} \right)$ is the inertia coefficient matrix, ${{D}} = diag\left( {\left[ {{D_1},{D_2},...,{D_{{N_g}}}} \right]} \right)$ is the damping coefficient matrix, $\bm{P_M} = {\left[ {{P_m}_1,{P_m}_2,...,{P_m}_{{N_g}}} \right]^T}$ is the vector of generators' mechanical power input, $\bm{P_E} = {\left[ {{P_e}_1,{P_e}_2,...,{P_e}_{{N_g}}} \right]^T}$ is the vector of generators' electromagnetic power output, ${{E_i}}$ is the constant voltage behind the transient reactance of the ${i}$th generator, ${G_{ij}}+{j}{B_{ij}}$ is the $\left( {i},{j} \right)$th entry of the reduced system admittance matrix with only generator buses, and $N_g$ is the number of generators.

Similar to \cite{load-gaussian-model,load-stochastics}, we make a common assumption that load active powers are perturbed by independent Gaussian noise from their base loadings. As previously shown in \cite{load-stochastics}, the load variations can be described by random perturbations at the diagonal elements of the reduced admittance matrix $Y\left( {i,i} \right) = {Y_{ii}}\left( {1 + {\sigma _i}{\xi _i}} \right)\angle {\phi _{ii}}$, where $i$ is the generator number, ${\xi _i}$ is a standard Gaussian variable and ${\sigma _i}^2$ describes the intensity of the fluctuations. Hence, the power system dynamic model considering the stochasticity of loads can be described by\cite{ref:wang2017pmu}:
\color{black}
\begin{equation}
\begin{aligned}
\mathop {\bm{\dot \delta }}   &= {\omega _0}\left( {{\bm{\omega }} - \bm{1}} \right)\\
{{M}}\mathop {\bm{\dot \omega }}   &= \bm{P_M} - \bm{P_E} - {{D}}\left( {{\bm{\omega }} - \bm{1}} \right) - {{{E}}^2}{G\Sigma}\bm{\xi}
\end{aligned} \label{eq2}
\end{equation}
\color{black}
where ${{E}} = diag\left( {\left[ {{E_1},{E_2},...,{E_{{N_g}}}} \right]} \right)$ , ${{G}} = diag\left( {\left[ {{G_{11}},{G_{22}},...,{G_{{N_g}{N_g}}}} \right]} \right)$ , ${{\Sigma }} = diag\left( {\left[ {{\sigma _1},{\sigma _2},...,{\sigma _{{N_g}}}} \right]} \right)$ and ${\bm{\xi }} =\left[ {{\xi _1},{\xi _2},...,{\xi _{{N_g}}}} \right]^T$. 
Linearizing (\ref {eq2}) around the steady-state operating point gives
\begin{equation}
\begin{array}{l}
{{\Delta }}\mathop {\bm{\dot \delta }}   = {\omega _0}{{\Delta \bm{\omega} }}\\
{{{M}}{\Delta }}\mathop {\bm{\dot \omega }}  =  -\Delta{\bm{P_E}}- {{{D}}{\Delta \bm{\omega} }} - {{{E}}^2}{G\Sigma}\bm{\xi}
\end{array} \label{eq3}
\end{equation}
which can be further represented in the compact form by substituting $\Delta{\bm{P_E}}=\frac{\partial{\bm{P_E}}}{\partial{\bm{\delta}}}{\Delta{\bm{\delta}}}$:
\begin{equation}
\mathop { \bm{\dot x}}   = {{{A}_{o}}\bm{x}} + {{{S}}\bm{\xi }} \label{eq4}
\end{equation}
where ${\bm{x}} = {\left[ {\Delta}\bm{{ \delta }},{\Delta}\bm{{ \omega }} \right]^T}$, ${{S}} = {\left[ {{{0}}, - {{{M}}^{ - 1}}{{{E}}^2}{{G\Sigma }}} \right]^T}$, ${{{A}}_{{o}}}=\left[\begin{array}{cc}{0}&\omega_0{I_{N_g}}\\
-M^{-1}{\frac{{\partial {{\bm{P}}_{\bm{E}}}}}{{\partial {\bm{\delta }}}}}&-M^{-1}D\end{array}\right]$ is the open-loop system state matrix, and ${{{I}}_{{N_g}}}$ is an identity matrix of size ${N_g}$. Particularly, $\bm{x}$ is a vector Ornstein-Uhlenbeck process, which is Gaussian and Markovian.
 
 $A_{o}$ carries significant information about the system dynamics and stability. For instance, the small-signal stability analysis of a power system is based on analyzing the eigenvalues and eigenvectors of $A_{o}$. Conventionally, $A_{o}$ can be easily calculated if it is assumed that the system dynamic model and the network topology are known, and the system states from the state estimator are accurate. Nevertheless, such assumption may not be true in practice due to network topology errors, bad data, etc. Therefore, an online data-driven method was proposed in \cite{ref:sheng2019online} to estimate the matrix ${{A}_{o}}$, which has been shown to be accurate, robust to measurement noise, and adaptive to topology change. However, the formulation in \cite{ref:sheng2019online} does not consider the presence of VSCs, which in turn is essential considering an increasing number of VSCs due to the growing penetration of RES. The integration of VSCs, on the other hand, makes the matrix estimation challenging, as the impacts of VSCs has to be carefully expressed in the formulation of (\ref{eq4}). An approach to separate the open-loop system state matrix $A$  and the input matrix $B$ that quantifies the impacts of VSCs is also needed for the sake of control design.  
 Details will be discussed in Section \ref{sectionmatrixestimation}. 
 
 Lastly, although the 2nd-order generator model is assumed in the proposed methodology, higher-order generator models with detailed control are used in the simulation study to show the feasibility of the proposed method in practice.

\subsection{VSC model}
Since we are interested in the electromechanical stability around the steady state, the VSCs are modelled as power sinks and sources, similar to the approach adopted in \cite{ref:multi-terminal-wind}, \color{black}which is based on the approximation that fast dynamics of VSCs are much faster than electromechanical dynamics such that VSCs respond instantaneously fast to the power reference change \cite{ref:preece2012probabilistic}.  

\color{black}
In order to provide damping support from VSCs, supplementary active and reactive power signals should be added to the steady state references \cite{ref:trinh2014methods}. Hence, {\color{black}the power injections 
from the buses where VSCs are connected} are
\begin{eqnarray} 
{{\bm{P}}_{\bm{v}}}& = &{{\bm{P}}_{{\bm{vs}}}} + {{\bm{P}}_{{\bm{vd}}}}\nonumber\\
{{\bm{Q}}_{\bm{v}}}& =& {{\bm{Q}}_{{\bm{vs}}}} + {{\bm{Q}}_{{\bm{vd}}}}\label{eq5}
\end{eqnarray}
where ${{\bm{P}}_{\bm{v}}} = {\left[ {{P_{v1}},{P_{v2}}, \cdots ,{P_{v{N_v}}}} \right]^T}$ represents the real-time active power references of the VSC; ${{\bm{P}}_{\bm{vs}}} = {\left[ {{P_{vs_1}},{P_{vs_2}} \cdots {P_{vs_{N_v}}}} \right]^T}$ denotes the steady-state active power references; ${{\bm{P}}_{\bm{vd}}} = {\left[ {{P_{vd_1}},{P_{vd_2}} \cdots {P_{vd_{N_v}}}} \right]^T}$ are the references for the supplementary active power generated by the damping controller. The reactive powers ${{\bm{Q}}_{\bm{v}}}$, ${{\bm{Q}}_{\bm{vs}}}$ and ${{\bm{Q}}_{\bm{vd}}}$ are defined in a similar way. \color{black}It should be noted that the intermittency of RES is not considered in this paper similar to \cite{ ref:mokhtari2014toward, ref:singh2014interarea}, because the time scale of RES intermittency (a few minutes \cite{ref:brouwer2014impacts}) is much larger than that of electromechanical dynamics (0.5s-5s \cite{ref:machowski2020power}).  In addition, \cite{ref:multi-terminal-wind} shows that the fluctuation of the output power of RES may be quickly smoothed out through proper power control (e.g., pitch angle control for wind turbines). \color{black}

In order to provide damping, the frequency variations of generators are  typically used as the feedback signals \cite{ref:multi-terminal-wind} and we have
\color{black}
\begin{equation}
\begin{array}{l}
{{\bm{P}}_{{\bm{vd}}}} = {{{K}}_{{1}}}\left( {{\bm{\omega }} - \bm{1}} \right)\\
{{\bm{Q}}_{{\bm{vd}}}} = {{K}_2}\left( {{\bm{\omega }} - \bm{1}} \right)
\end{array} \label{eq6}
\end{equation}
\color{black}
where ${{K}_{{1}}}$ and ${{K}_{{2}}}$ are damping coefficients of the active and reactive power control of VSCs, respectively.

Linearizing (\ref {eq5}) and (\ref {eq6}) around the steady state gives
\begin{equation}
\begin{array}{l}
{{\Delta }}{{\bm{P}}_{\bm{v}}} = {{\Delta }}{{\bm{P}}_{{\bm{vd}}}} = {{K}_{{1}}}{{\Delta \bm{\omega} }}\\
{{\Delta }}{{\bm{Q}}_{\bm{v}}} = {{\Delta }}{{\bm{Q}}_{{\bm{vd}}}} = {{K}_{{2}}}{{\Delta \bm{\omega} }}
\end{array} \label{eq7}
\end{equation}

\subsection{Integration of VSCs into AC system}

\begin{figure} \centering
  \includegraphics[width=0.6\linewidth]{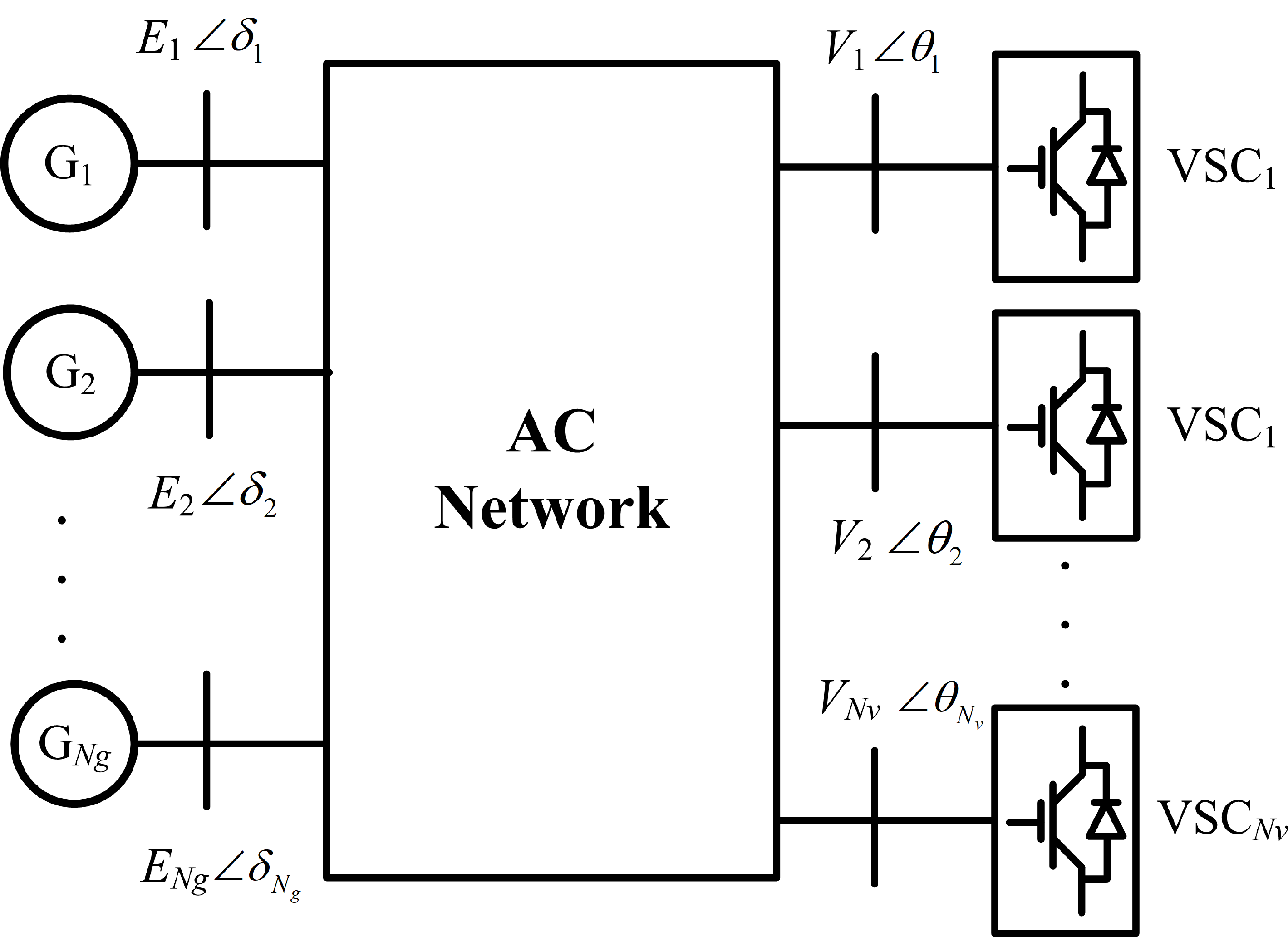}
  \caption{An AC system integrated with VSCs.}
  \label{fig1}
\end{figure}

By applying Kron reduction \cite{kron-reduction}, we can eliminate all buses except the ${N_g}$  generator buses and the ${N_v}$ VSC buses as shown in Fig.~\ref{fig1}. The {\color{black}active power injection from the bus where the $i$th generator is connected} can be calculated by 
\begin{equation}
\begin{array}{l}
{P_{ei}} = \sum\limits_{j = 1}^{{N_g}} {{E_i}{E_j}\left( {{G_{GGij}}\cos \left( {{\delta _i} - {\delta _j}} \right) + {B_{GGij}}\sin \left( {{\delta _i} - {\delta _j}} \right)} \right)} \\
 + \sum\limits_{j = 1}^{{N_v}} {{E_i}{V_j}\left( {{G_{GVij}}\cos \left( {{\delta _i} - {\theta _j}} \right) + {B_{GVij}}\sin \left( {{\delta _i} - {\theta _j}} \right)} \right)} 
\end{array}\label{eq8}
\end{equation}
where ${V_j}$ and ${\theta _j}$ are the voltage magnitude and voltage angle of VSC bus ${j}$. ${G_{GGij}}$ and ${B_{GGij}}$ are the equivalent conductance and susceptance between generator buses ${i}$ and ${j}$. ${G_{GVij}}$ and ${B_{GVij}}$ are the equivalent conductance and susceptance between generator bus ${i}$ and VSC bus ${j}$.

Similarly, the {\color{black}active and reactive power injections 
from the bus where the $i$th VSC is connected} can be expressed by
\begin{equation}
\begin{array}{l}
{P_{vi}} = \sum\limits_{j = 1}^{{N_g}} {{V_i}{E_j}\left( {{G_{VGij}}\cos \left( {{\theta _i} - {\delta _j}} \right) + {B_{VGij}}\sin \left( {{\theta _i} - {\delta _j}} \right)} \right)} \\
 + \sum\limits_{j = 1}^{{N_v}} {{V_i}{V_j}\left( {{G_{VVij}}\cos \left( {{\theta _i} - {\theta _j}} \right) + {B_{VVij}}\sin \left( {{\theta _i} - {\theta _j}} \right)} \right)} \\
{Q_{vi}} = \sum\limits_{j = 1}^{{N_g}} {{V_i}{E_j}\left( {{G_{VGij}}\sin \left( {{\theta _i} - {\delta _j}} \right) - {B_{VGij}}\cos \left( {{\theta _i} - {\delta _j}} \right)} \right)} \\
 + \sum\limits_{j = 1}^{{N_v}} {{V_i}{V_j}\left( {{G_{VVij}}\sin \left( {{\theta _i} - {\theta _j}} \right) - {B_{VVij}}\cos \left( {{\theta _i} - {\theta _j}} \right)} \right)} 
\end{array}\label{eq9}
\end{equation}
where ${G_{VGij}}$ and ${B_{VGij}}$ are the equivalent conductance and susceptance between VSC bus ${i}$ and generator bus ${j}$. ${G_{VVij}}$ and ${B_{VVij}}$ are the equivalent conductance and susceptance between VSC bus ${i}$ and VSC bus ${j}$.

Linearizing the power injections of generators and VSCs (\ref{eq8})-(\ref{eq9}) around the steady state yields
\begin{equation}
\left[ \begin{array}{l}
{{\Delta }}{\bm{P_E}}\\
{{\Delta }}{{\bm{P}}_{\bm{v}}}\\
{{\Delta }}{{\bm{Q}}_{\bm{v}}}
\end{array} \right]{\bm{ = }}\left[ \begin{array}{l}
\frac{{\partial {\bm{P_E}}}}{{\partial {\bm{\delta }}}}{\kern 1pt} {\kern 1pt} {\kern 1pt} {\kern 1pt} {\kern 1pt} {\kern 1pt} {\kern 1pt} {\kern 1pt} \frac{{\partial {\bm{P_E}}}}{{\partial {\bm{\theta }}}}{\kern 1pt} {\kern 1pt} {\kern 1pt} {\kern 1pt} {\kern 1pt} {\kern 1pt} {\kern 1pt} {\kern 1pt} \frac{{\partial {\bm{P_E}}}}{{\partial {\bm{V}}}}\\
\frac{{\partial {{\bm{P}}_{\bm{v}}}}}{{\partial {\bm{\delta }}}}{\kern 1pt} {\kern 1pt} {\kern 1pt} {\kern 1pt} {\kern 1pt} {\kern 1pt} {\kern 1pt} {\kern 1pt} {\kern 1pt} \frac{{\partial {{\bm{P}}_{\bm{v}}}}}{{\partial {\bm{\theta }}}}{\kern 1pt} {\kern 1pt} {\kern 1pt} {\kern 1pt} {\kern 1pt} {\kern 1pt} {\kern 1pt} {\kern 1pt} \frac{{\partial {{\bm{P}}_{\bm{v}}}}}{{\partial {\bm{V}}}}\\
\frac{{\partial {{\bm{Q}}_{\bm{v}}}}}{{\partial {\bm{\delta }}}}{\kern 1pt} {\kern 1pt} {\kern 1pt} {\kern 1pt} {\kern 1pt} {\kern 1pt} \frac{{\partial {{\bm{Q}}_{\bm{v}}}}}{{\partial {\bm{\theta }}}}{\kern 1pt} {\kern 1pt} {\kern 1pt} {\kern 1pt} {\kern 1pt} {\kern 1pt} \frac{{\partial {{\bm{Q}}_{\bm{v}}}}}{{\partial {\bm{V}}}}
\end{array} \right]\left[ \begin{array}{l}
{{\Delta \bm{\delta} }}\\
{{\Delta \bm{\theta} }}\\
{{\Delta \bm{V}}}
\end{array} \right]\label{eq10}
\end{equation}

Since the system is assumed to be in the normal operating state such that the Jacobian matrix is well-conditioned \cite{well-conditioned}, we can represent ${{\Delta }}{\bm{P_E}}$ by ${{\Delta \bm{\delta} }}$,  $\Delta \bm{P_v}$ and $\Delta \bm{Q_v}$: 
\begin{equation}
{{\Delta }}{\bm{P_E}} = {{{A}}_{{1}}}{{\Delta \bm{\delta} }} + {{{A}}_{{2}}}{{\Delta }}{{\bm{P}}_{\bm{v}}} + {{{A}}_{{3}}}{{\Delta }}{{\bm{Q}}_{\bm{v}}} \label{eq11}
\end{equation}
The detailed expression of ${A_1}$, ${A_2}$ and ${A_3}$ can be found in Appendix \ref{Aderivation}. Substituting the expression of ${{\Delta }}{\bm{P_E}}$ from (\ref{eq11}) to (\ref{eq3}) leads to
\begin{equation}
\begin{aligned}
\left[ \begin{array}{l}
{{\Delta }}\mathop {\bm{\dot \delta }}  \\
{{\Delta }}\mathop {\bm{\dot \omega }}  
\end{array} \right] &= \left[ \begin{array}{cc}
{0}&{\omega _0}{{{I}}_{{N_g}}}\\
 - {{{M}}^{ - 1}}{{{A}}_{{1}}}&- {{{M}}^{ - 1}}{{D}}
\end{array} \right]\left[ \begin{array}{l}
{{\Delta \bm{\delta} }}\\
{{\Delta \bm{\omega} }}
\end{array} \right]\\
 &+ \left[ \begin{array}{cc}
{0}&{0}\\
- {{{M}}^{ - 1}}{{{A}}_{{2}}}&- {{{M}}^{ - 1}}{{{A}}_{{3}}}
\end{array} \right]\left[ \begin{array}{l}
{{\Delta }}{{\bm{P}}_{\bm{v}}}\\
{{\Delta }}{{\bm{Q}}_{\bm{v}}}
\end{array} \right] \\
&+ \left[ \begin{array}{c}
{0}\\
 - {{{M}}^{ - 1}}{{{E}}^2}{{G\Sigma }}
\end{array} \right]{\bm{\xi }}
\end{aligned}  \label{eq12}
\end{equation}

It can be further represented in the compact form:
\begin{equation}
\mathop { \bm{\dot x}}   = {{{A}}\bm{x}} + {{{B}}\bm{u}} + {{{S}}\bm{\xi }} 
\label{eq13}
\end{equation}
 where ${\bm{u}} = {\left[ {{{\Delta }}{{\bm{P}}_{\bm{v}}},{{\Delta }}{{\bm{Q}}_{\bm{v}}}} \right]^T}$, ${{A}} =\left[\begin{array}{l}    {\kern 1pt} {\kern 1pt} {\kern 1pt} {\kern 1pt} {{0}}{\kern 1pt} {\kern 1pt} {\kern 1pt} {\kern 1pt} {\kern 1pt} {\kern 1pt} {\kern 1pt} {\kern 1pt} {\kern 1pt} {\kern 1pt} {\kern 1pt} {\kern 1pt} {\kern 1pt} {\kern 1pt} {\kern 1pt} {\kern 1pt} {\kern 1pt} {\kern 1pt} {\kern 1pt} {\kern 1pt} {\kern 1pt} {\kern 1pt} {\kern 1pt} {\kern 1pt} {\kern 1pt} {\omega _0}{{{I}}_{{N_g}}}\\
 \bar{{{A}}_{{1}}}{\kern 1pt} {\kern 1pt} {\kern 1pt} {\kern 1pt} {\kern 1pt} {\kern 1pt} {\kern 1pt} {\kern 1pt}  - {{{M}}^{ - 1}}{{D}} \end{array} \right]$, ${{B}} =\left[\begin{array}{l}   {\kern 1pt} {\kern 1pt} {{0}}{\kern 1pt} {\kern 1pt} {\kern 1pt} {\kern 1pt} {\kern 1pt} {\kern 1pt} {\kern 1pt} {\kern 1pt} {\kern 1pt} {\kern 1pt} {\kern 1pt} {\kern 1pt} {\kern 1pt} {\kern 1pt} {\kern 1pt} {\kern 1pt} {\kern 1pt} {\kern 1pt} {\kern 1pt} {\kern 1pt} {\kern 1pt} {{0}}\\
 \bar{{{A}}_{{2}}}{\kern 1pt} {\kern 1pt} {\kern 1pt} {\kern 1pt} {\kern 1pt} {\kern 1pt} {\kern 1pt} {\kern 1pt} {\kern 1pt} {\kern 1pt} {\kern 1pt} {\kern 1pt} \bar{{{A}}_{{3}}} \end{array} \right]$. Particularly, $\bar {{{{A}}_{{1}}}{\kern 1pt} }  =  - {{{M}}^{ - 1}}{{{A}}_{{1}}}$, $\bar {{{{A}}_{{2}}}}  =  - {{{M}}^{ - 1}}{{{A}}_{{2}}}$ and $\bar {{\kern 1pt} {{{A}}_{{3}}}} {\kern 1pt} {\kern 1pt}  =  - {{{M}}^{ - 1}}{{{A}}_{{3}}}$.

In order to design the damping controller, the accurate information of ${A}$ and ${B}$ in (\ref{eq13}) needs to be known. Substituting (\ref{eq7}) into (\ref{eq13}), $B$ can be embedded in the system state matrix as follows: 
\begin{equation}
\mathop {\bm{\dot x}}   = {{{A}}_{{c}}}{\bm{x}} + {{S\bm{\xi} }} 
\label{eq14}
\end{equation}
where 
${{{A}}_{{c}}}=\left[\begin{array}{cc}{0}&\omega_0{I_{N_g}}\\
\bar{A_1}&\bar{A_2}{K_1}+\bar{A_3}{K_2}-M^{-1}D\end{array}
\right]$.

Compared with the open-loop matrix ${{A}_{o}}$ in (\ref{eq4}) without VSC, the closed-loop system state matrix ${{A}_{c}}$ includes the impact of VSCs, as reflected by the new elements $\bar{A_2}{K_1}$ and $\bar{A_3}{K_2}$. Since the work in \cite{ref:trinh2014methods, ref:LQG} has shown that the impact of reactive power on damping control is much smaller than that of active power, 
the reactive power control for damping is not considered in the rest of paper, i.e., ${{\Delta }}{{\bm{Q}}_{\bm{v}}}$ is zero by setting ${K_2}$ to be zero matrix. As a result, $A$ remains the same as in (\ref{eq13}):
\begin{equation}
    {{A}} =\left[\begin{array}{cc} {0}&{\omega _0}{{{I}}_{{N_g}}}\\
 \bar{{{A}}_{{1}}}&- {{{M}}^{ - 1}}{{D}} \end{array} \right]\label{eq:A}
\end{equation}
${B}$ becomes:
\begin{equation}
 {{B}} =\left[\begin{array}{c} {{0}}\\
 \bar{{{A}}_{{2}}}\end{array} \right]\label{eq:B}
\end{equation}
$A_c$ becomes:
\begin{equation}
{{{A}}_{{c}}}=\left[\begin{array}{cc}{0}&\omega_0{I_{N_g}}\\
\bar{A_1}&\bar{A_2}{K_1}-M^{-1}D\end{array}
\right]\label{eq:Acl}
\end{equation}

\section{The proposed WAMS-based method for estimating matrices} \label{sectionmatrixestimation}

The WAMS-based damping control design is comprised of two steps: estimating the matrices $A$ and $B$ in the current operating condition; designing the damping coefficients ${K_1}$ of VSCs based on the estimated $A$ and $B$ for a desired damping performance.

\subsection{The Theoretical Basis of the WAMS-Based Method for Matrix Estimation}
In the compact form of the power system dynamic model incorporating VSCs (\ref{eq14}), $\bm{x}$ is a multivariate Ornstein-Uhlenbeck process. According to the regression theorem of a multivariate Ornstein-Uhlenbeck process \cite{regression-theorem}, if the dynamic system described by (\ref{eq14}) is stable which is typically satisfied if the system is in the normal operating condition, the $\tau$-lag time correlation matrix $R(\tau)$ satisfies the following differential equation: 
\begin{equation}
    \frac{d}{{d\tau }}\left[ {{{R}}\left( \tau  \right)} \right] =   {{{A}}_{{c}}}{{R}}\left( \tau  \right)
    \label{eq15}
\end{equation}
where ${{R}}\left( \tau  \right) \triangleq \left\langle {\left[ {{\bm{x}}\left( {t + \tau } \right) - \bar {\bm{x}} } \right], {{\left[ {{\bm{x}}\left( t \right) - \bar {\bm{x}} } \right]}^T}} \right\rangle $, 
and $\bm{\bar{x}}$ denotes the mean of $\bm{x}$. \color{black} The $\tau$-lag time correlation matrix describes the correlation of a random vector with itself at time lag $\tau$. 
\color{black}
Therefore, the system state matrix can be obtained by solving (\ref{eq15}) according to \cite{ref:sheng2019online} 
\begin{equation}
    {{{A}}_{{c}}} = \frac{1}{\tau }\log \left[ {{{R}}\left( \tau  \right){{{C}}^{ - 1}}} \right]
    \label{eq:estA_c}
\end{equation}
where the stationary covariance matrix  ${{C}} \triangleq \left\langle {\left[ {{\bm{x}}\left( t \right) - \bar {\bm{x}} } \right], {{\left[ {{\bm{x}}\left( t \right) - \bar {\bm{x}} } \right]}^T}} \right\rangle=R(0)$. 

Note that the statistics $R(\tau)$ and $C$ of 
{{$\bm{x}=[\Delta\bm{\delta}, \Delta\bm{\omega}]^T$}} can be estimated from PMU measurements ({see Appendix \ref{Approximation}} \color{black}). Equation (\ref{eq:estA_c}) indicates an ingenious way of estimating the physical model knowledge $A_{c}$ from the statistical properties of PMU measurements, which serves as the theoretical basis of the proposed method for estimating matrices. 

\subsection{The Proposed Algorithm for Estimating Matrices}
In order to estimate $A$ and $B$ in (\ref{eq:A})-(\ref{eq:B}), we propose an algorithm to estimate the dynamic components $-M^{-1}D$, $\bar{A_1}$, and $\bar{A_2}$, respectively. We assume that all generator buses are equipped with PMUs so that generators' rotor angles and speeds in ambient conditions can be calculated using the approaches in \cite{zhou2014PMU, yan2011PMU}. The equipment of PMUs at all generators might seem optimistic for now, yet is reasonable in the near future because of the wide adoption of PMUs worldwide {\cite{ref:chen2015measurement}}. \color{black} In addition, we assume that 
VSCs are within their capability limits to provide damping and have enough controllability over the critical modes. 
\color{black} 
Also, the current damping coefficients ${K_1}$ of VSCs in (\ref{eq:Acl}) are assumed to be known when the algorithm is initiated (i.e., the default setting or no support). The specific steps of the proposed algorithm are as follows:

\textbf{Step 1}: Given PMU measurements with a sufficient window length, estimate $[\Delta\bm{\delta}, \Delta\bm{\omega}]^T$ and compute their sample covariance matrix $\hat{C}$ and sample $\tau$-lag correlation matrix $\hat{R}$ {\color{black}for a selected $\tau$ according to (\ref{Approximation of matrix 1}) and (\ref{Approximation of matrix 2}) in Appendix \ref{Approximation}}. Estimate the closed-loop system state matrix by (\ref{eq:estA_c}). 
\begin{equation}
{{\hat{A}}_{{c1}}} = \left[ \begin{array}{cc}
{{0}}& {\omega _0}{{{I}}_{{N_g}}}\\
{{{A}}_{{{c}}1{{LL}}}}& {{{A}}_{{{c}}1{{LR}}}}
\end{array} \right]  =\left[\begin{array}{cc}{0}&\omega_0{I_{N_g}}\\
\bar{A_1}&\bar{A_2}{K_1}-M^{-1}D\end{array}
\right] \label{eq18}
\end{equation}

\textbf{Step 2}: Add  small perturbations (e.g., $\alpha\%$) to the damping coefficients of all VSCs simultaneously by ${{K}_{{1}}}( {i,j})\leftarrow{{K}_{{1}}}( {i,j})+{\Delta{K}_{{1}}}( {i,j})$, where $K_1\in \mathbb{R}^{N_v\times N_g}$, 
\begin{equation}
{\Delta{K}_{{1}}}\left( {i,j} \right) = \left\{ \begin{array}{l}
\alpha\% {{K}_{{1}}}\left( {i,j} \right),{\kern 1pt} {\kern 1pt} {\kern 1pt}j = {g_i}\\
0,{\kern 1pt} {\kern 1pt} {\kern 1pt}j \ne {g_i}
\end{array} \right. \label{eq20}
\end{equation}
where ${g_i}$ represents the column of the entry that has the largest absolute value among $\{1,2,...,N_g\}\setminus\{g_1,...g_{i-1}\}$ columns in the ${i}$th row {\color{black} of $K_1$}. 

{\color{black} More specifically, $g_i$ is determined by the following steps.} {Starting} from the first VSC (e.g., 1st row of $K_1$), let $g_1$ be the column number that has the largest absolute value in the $1$st row; go to the $2$nd VSC (i.e., $2$nd row of $K_1$) and let $g_2$ be the column number that has the largest absolute value among $\{1,2,...,N_g\}\setminus\{g_1\}$ columns in the $2$nd row. We continue the above procedure until the last row of $K_1$. As such, the perturbation matrix $\triangle K_1$ will be a generalized permutation matrix such that each column of $\bar{A}_2$ will be detectable in the estimation (see (\ref{eq:A2est})). 

\textbf{Step 3}: Collect new PMU measurements and estimate $[\Delta\bm{\delta}, \Delta\bm{\omega}]^T$ after the perturbation.  Estimate the new system state matrix after the perturbation by (\ref{eq:estA_c}): 
\begin{equation}
\begin{aligned}
{{\hat{A}}_{{c2}}} &= \left[ \begin{array}{cc}
{{0}}& {\omega _0}{{{I}}_{{N_g}}}\\
{{{A}}_{{{c}}2{{LL}}}}& {{{A}}_{{{c}}2{{LR}}}}
\end{array} \right]  \\
&=\left[\begin{array}{cc}{0}&\omega_0{I_{N_g}}\\
\bar{A_1}&\bar{A_2}({K_1}+{\Delta{K}_{{1}}})-M^{-1}D\end{array}
\right] \label{eq22}
\end{aligned}
\end{equation}

\textbf{Step 4:} Estimate the dynamic components ${\bar{A_1}}$, $\bar {{{{A}}_{{2}}}{\kern 1pt} }$ and $-{M^{-1}D}$ as follows:
\begin{eqnarray}
\bar{A_1}^{est} &=& {A}_{{c}2{LL}} \label{eq19}\\
{\bar {{{{A}}_2}}}^{est}  &=& \left( {{{{A}}_{{{c}}2{{LR}}}} - {{{A}}_{{{c}}1{{LR}}}}} \right)\Delta{K}_{{1}}^{+}\label{eq:A2est}\\
({{-{M^{-1}D}}})^{est}&=&A_{c1LR}-\bar{A_2}^{est}{K_1}  \label{eq:MDest}
\end{eqnarray}
where ${\Delta{K}_{{1}}}^{+}$ is the pseudo inverse matrix of ${\Delta{K}_{{1}}}$.

Particularly, (\ref{eq:A2est}) is obtained by subtracting (\ref{eq18}) from (\ref{eq22}) {using a} simple manipulation to the lower-right part. Equation (\ref{eq:MDest}) is obtained by substituting $\bar{A_2}^{est}$ in (\ref{eq:A2est}) back to the lower-right part of (\ref{eq18}).

\noindent\textit{Remarks:}\\
\begin{itemize}
\item  In this paper, a window size of 300s is used in \textbf{Step 1} and \textbf{Step 3}, respectively, for which a good accuracy is achieved. It should be noted that, despite of a relatively large window size,  the above algorithm does not assume any model information (e.g. network topology, generator dynamic parameters $D$ and $M$), but can estimate the dynamic components as well as the matrices $A$ and $B$ purely from PMU data. Therefore, in the case of incomplete or inaccurate network information, the proposed method provides a way to acquire the dynamical system model for monitoring or control if sufficient data is available. 
\item {\color{black}The way to choose $g_i$ in \textbf{Step 2} is not unique. Different $g_i$ can be selected as long as the resulting $\triangle K_1$ is a generalized permutation matrix such that each column of $\bar{A}_2$ is detectable in the estimation. Regarding the feasibility of adding perturbations to VSCs simultaneously, it can be realized by using the Global Positioning System (GPS) satellite and the high speed communication network between the control center and distributed VSCs. A more detailed implementation in the real systems can be found in \cite{ref:taylor2005wacs, ref:peng2009implementation}.}
\item  The second-order generator model is assumed in the proposed algorithm. However, 3rd-order generator models with automatic voltage regulators (AVR) are used in the simulation study to show the effectiveness of the algorithm in practical applications.
\end{itemize}

\section{The WAMS-based method for wide-area damping control}

Once the dynamic components $ - {{{M}}^{ - 1}}{{D}}$, $\bar{A}_1$ and $\bar{A}_2$ are estimated, ${{A}}$ and ${{B}}$ in (\ref{eq13}) can be obtained from (\ref{eq:A})-(\ref{eq:B}). Therefore, the control input ${\bm{u}}$ can be designed to provide effective damping. To this end, the MLQR \cite{ref:almutairi2010enhancement} is applied to design the system input $\bm{u}$ in (\ref{eq13}) such that the following quadratic cost function is minimized:
\color{black}
\begin{equation}
J_C=\mathop {\lim }\limits_{t \to \infty } E\left\{ {\int\limits_0^t {\left( {{{\bm{x}}^T}({{{L}}^T}{{{W}}_{{Q}}}{{L})\bm{x}} + {{\bm{u}}^T}{{{W}}_{{R}}}{\bm{u}}} \right)dt} } \right\}
\label{eq:costfunc}
\end{equation}
\color{black}
where ${{{W}}_{{Q}}}\geq0$ and ${{{W}}_{{R}}}>0$ are weighting matrices which in most cases are set as diagonal matrices. In specific, higher diagonal values in ${{{W}}_{{Q}}}$ correspond to a greater desire to stabilize the corresponding oscillation modes. Higher diagonal values in ${{{W}}_{{R}}}$ represent a more strict requirement to reduce the corresponding control inputs. Note that only the relative sizes of the components in the weighting matrices matter rather than the absolute values. Besides, ${{L}}$ is the mapping matrix obtained from the Real Schur Decomposition \cite{mapping-matrix} {\color{black}of the system state matrix $A$}, which transforms state variables ${\bm{x}}\left( t \right)$ to the modal variables ${\bm{z}}\left( t \right)$: 
\begin{equation}
{\bm{z}}\left( t \right) = {{L}\bm{x}}\left( t \right) \label{eq26}
\end{equation}
where $\bm{z}$ are directly associated with system
modes $e^{\lambda_i t}$, $i=1,2,...,2\times N_g$. As a result, to damp the critical modes, we can add weights only to the corresponding diagonal values of $W_Q$ while setting all the other values to be zero, making the well-damped modes unaffected by the control. 

The MLQR controller gain $\Gamma$ can be obtained by solving the associated algebraic Riccati equation (ARE) according to the cost function (\ref{eq:costfunc}). The final MLQR feedback control law is:
\begin{equation}
    \bm{u}=-\Gamma\bm{x}
\end{equation}
The damping coefficients ${K_1}$ of VSCs, therefore, are set according to $\Gamma$. 

The main advantage of the MLQR control method lies in the fact that a full decoupling between modes can be reached such that multiple critical modes can be damped simultaneously while well-damped modes remain unaffected, making the local controllers effective \cite{ref:almutairi2010enhancement}. In addition, the VSCs' modulation capacities can be incorporated by tuning the weighting matrix $W_R$.

\begin{figure} \centering
  \includegraphics[width=0.8\linewidth]{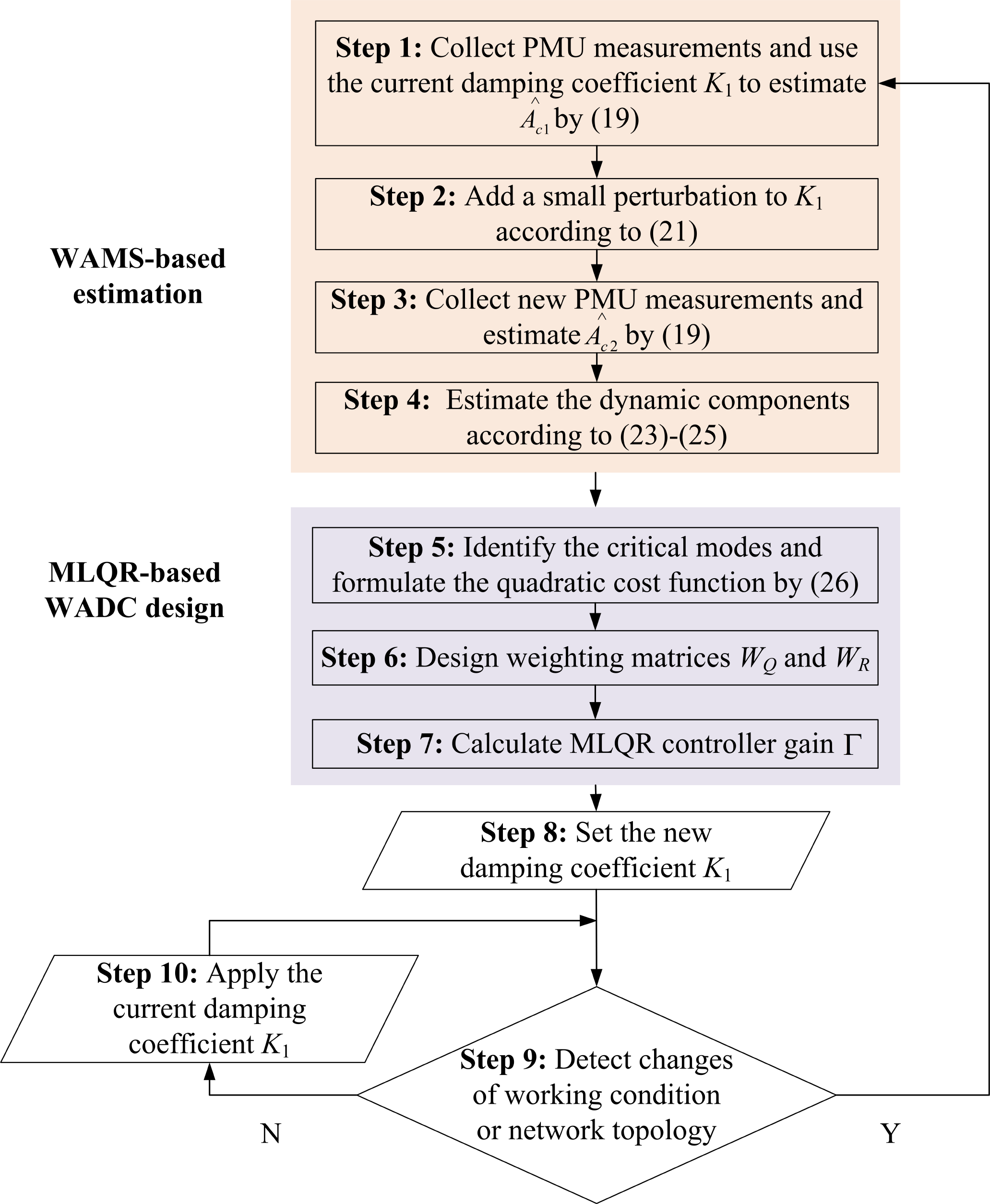}
  \caption{Flow chart of the WAMS-based WADC strategy.}
  \label{Flow chart} 
\end{figure}

To sum up, the overall procedure for the online WAMS-based WADC method using VSCs is described in Fig.~\ref{Flow chart}. It consists of two stages: the WAMS-based method for estimating matrices and the MLQR-based WADC design. The damping coefficients can be adjusted if the changes of working condition or network topology are detected. \color{black} It should be noteworthy that if enough damping is provided by VSCs when the system works in the normal working condition, the system may still have suboptimal damping performance for the critical modes within the small time span after the event, when the controller is not updated yet. In the extreme case where very poor damping occurs, some emergency control strategy (e.g., PSS-based) using ring-down PMU data \cite{ref:pradhan2018model} can be incorporated. \color{black}

\section{Case Studies}

\begin{figure} \centering
  \includegraphics[width=0.7\linewidth]{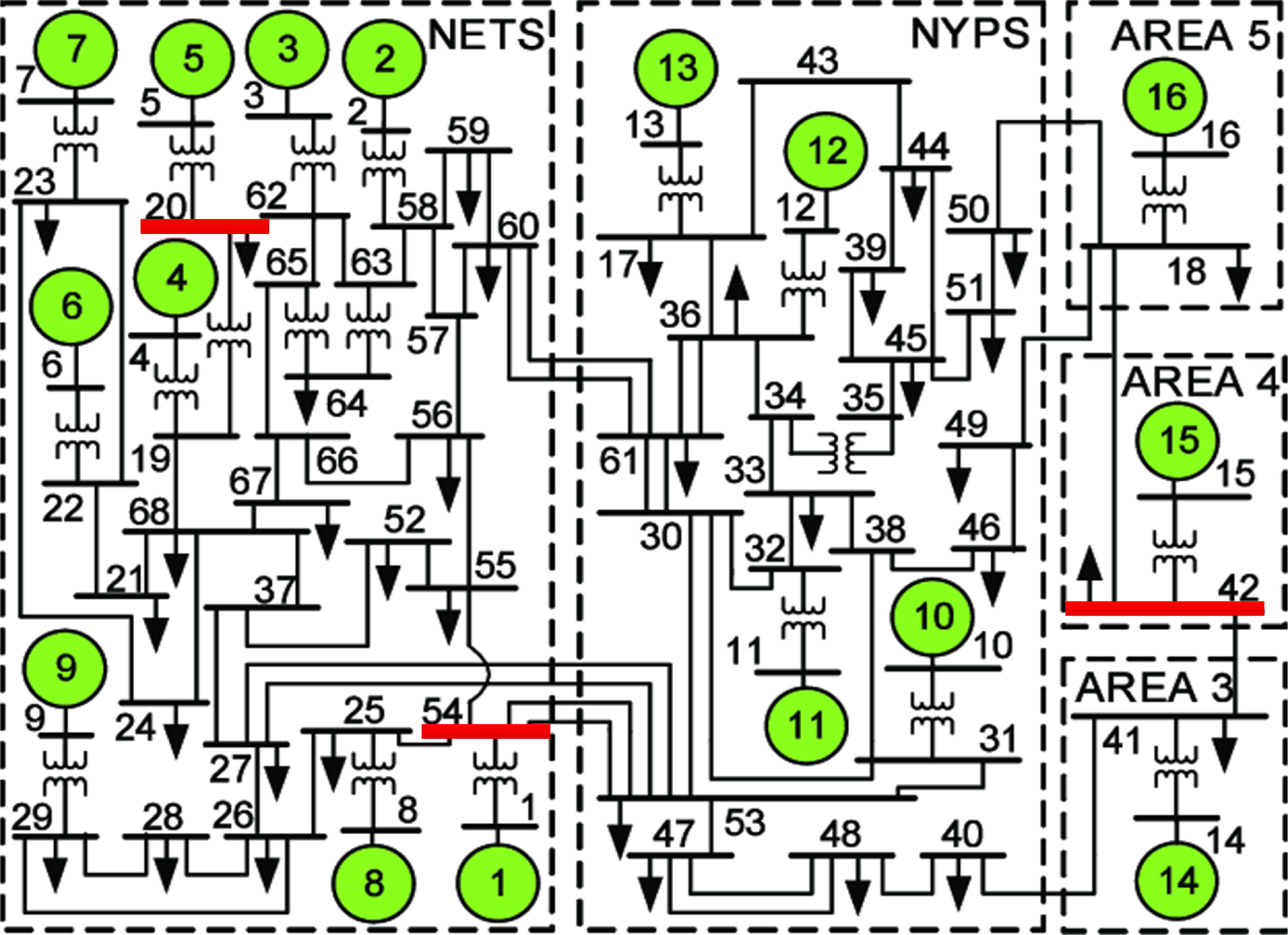}
  \caption{The network topology of the IEEE 68-bus system with integrated VSCs.}
  \label{network topology}
\end{figure}

The IEEE 68-bus system is modified to test the proposed WAMS-based WADC method. In particular, {three VSCs (denoted by VSC1, VSC2 and VSC3) are placed at bus 54, 20 and 42 respectively, which are marked in red in Fig.~\ref{network topology}.} \color{black} The implementation for different VSC locations will be discussed in Section \ref{section_diffVSCs}. \color{black} In steady state, all VSCs inject the same amount of active power of 0.5 p.u. into the AC system and there are no reactive power exchanges. In order to validate the effectiveness of the proposed method in practical applications, the 3rd-order generator models are used throughout the simulations. 
In addition, G1-G12 are controlled by automatic voltage regulators (AVRs). The fluctuation intensities ${\sigma _1},...,{\sigma _n}$ describing load variations are all set to 0.05. \color{black} $\tau=100$ ms is used to calculate the correlation matrix (see (\ref{Approximation of matrix 2})). \color{black} All the simulations are done in the Power System Analysis Toolbox (PSAT) \cite{ref:milano2005open}. 

\begin{figure} \centering
  \includegraphics[width=0.7\linewidth]{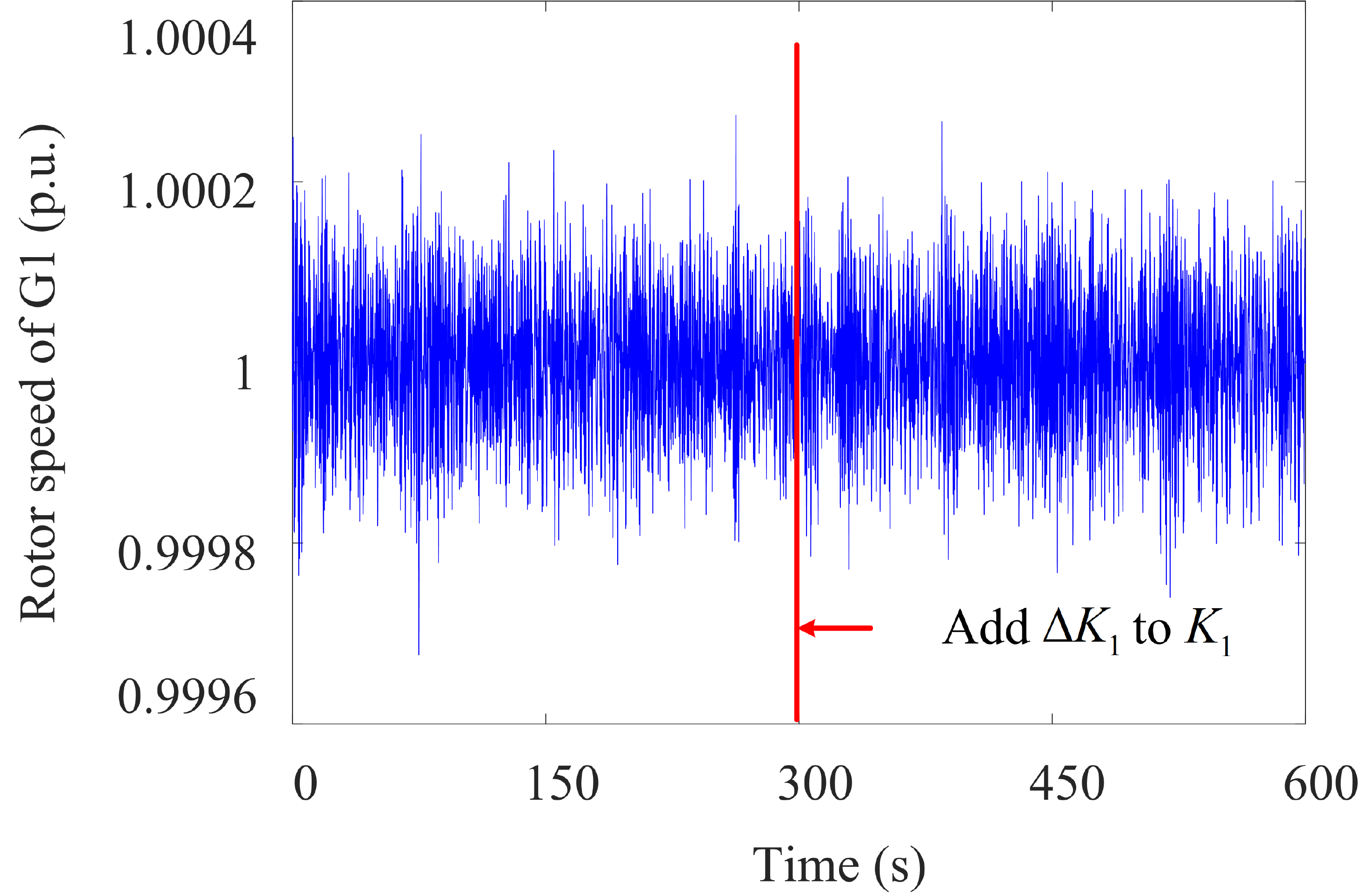}
  \caption{Trajectories of G1's rotor speed before and after adding the small perturbation.}
  \label{stochastics}
\end{figure}

\subsection{Validation For the Algorithm of Estimating Matrices}

\begin{figure} [htbp]
\centering
\subfigure[Values of $ {{{M}}^{ - 1}}{{D}}$]{ \label{fig6a} 
\begin{minipage}[t]{0.5\linewidth}
\centering
\includegraphics[width=1.8in]{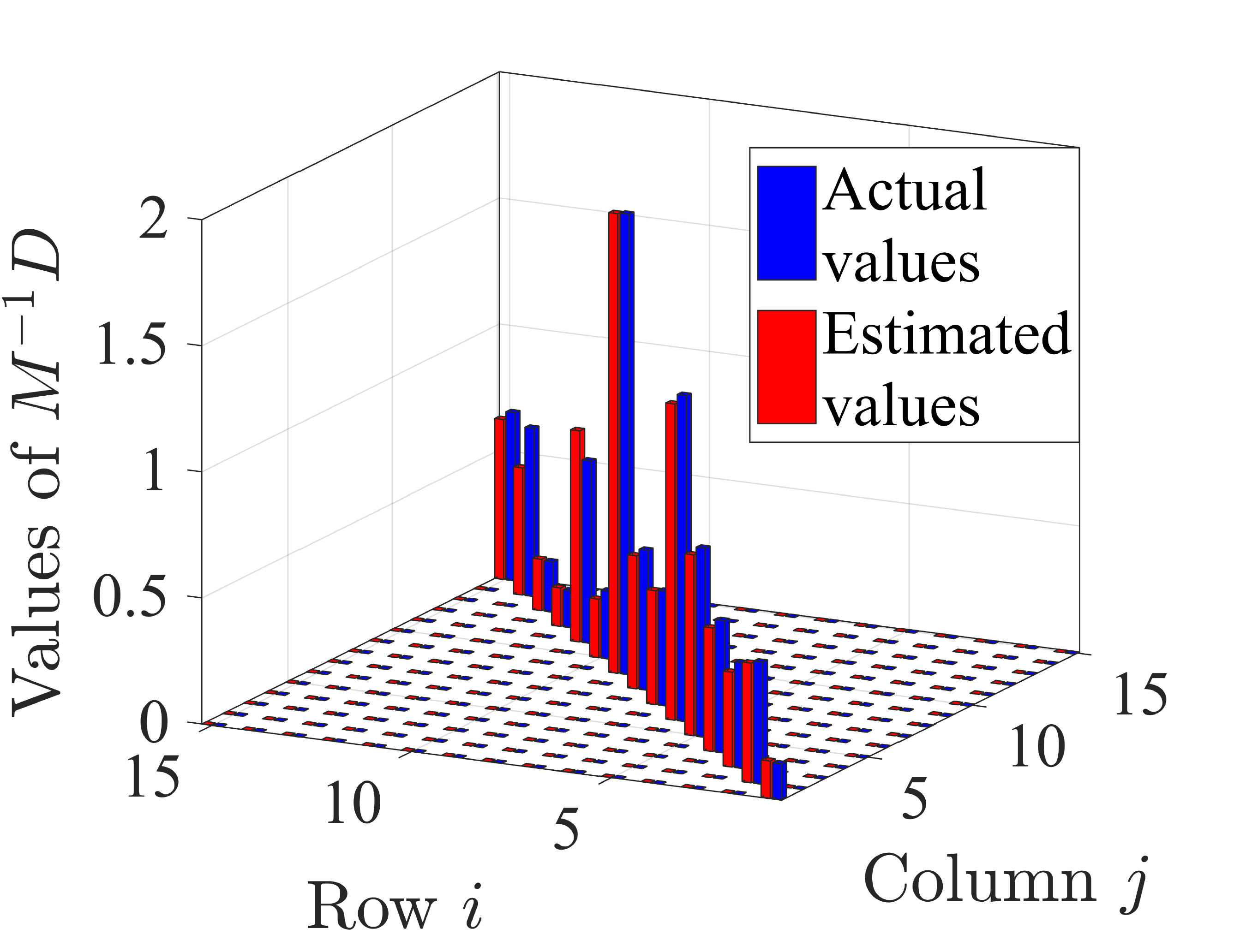}
\end{minipage}%
}%

\subfigure[Values of $-\bar {{{{A}}_{{1}}}{\kern 1pt} }$]{ \label{fig6b} 
\begin{minipage}[t]{0.5\linewidth}
\centering
\includegraphics[width=1.8in]{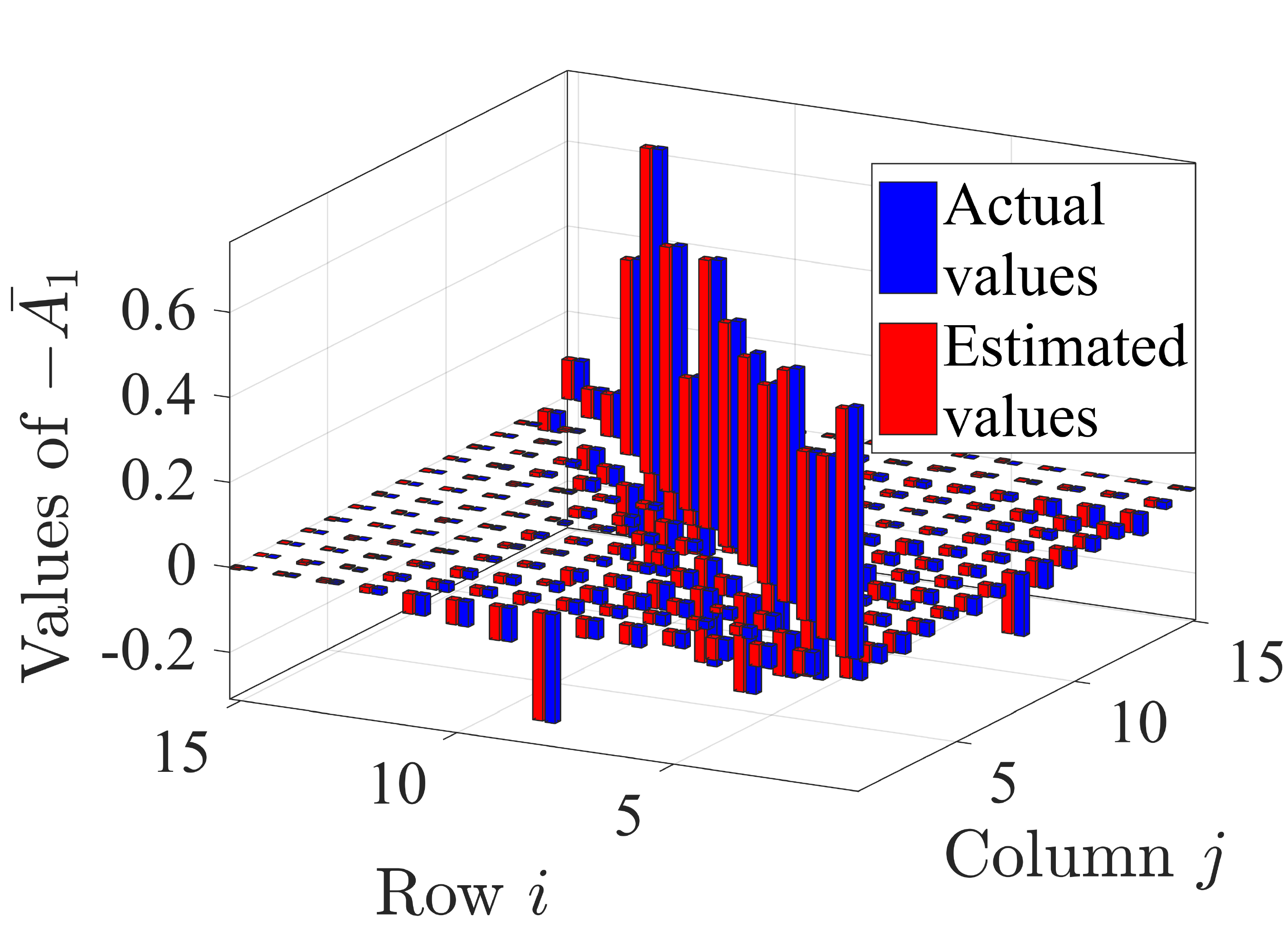}
\end{minipage}%
}%
\subfigure[Values of $\bar {{{{A}}_{{2}}}{\kern 1pt} }$]{ \label{fig6c} 
\begin{minipage}[t]{0.5\linewidth}
\centering
\includegraphics[width=1.8in]{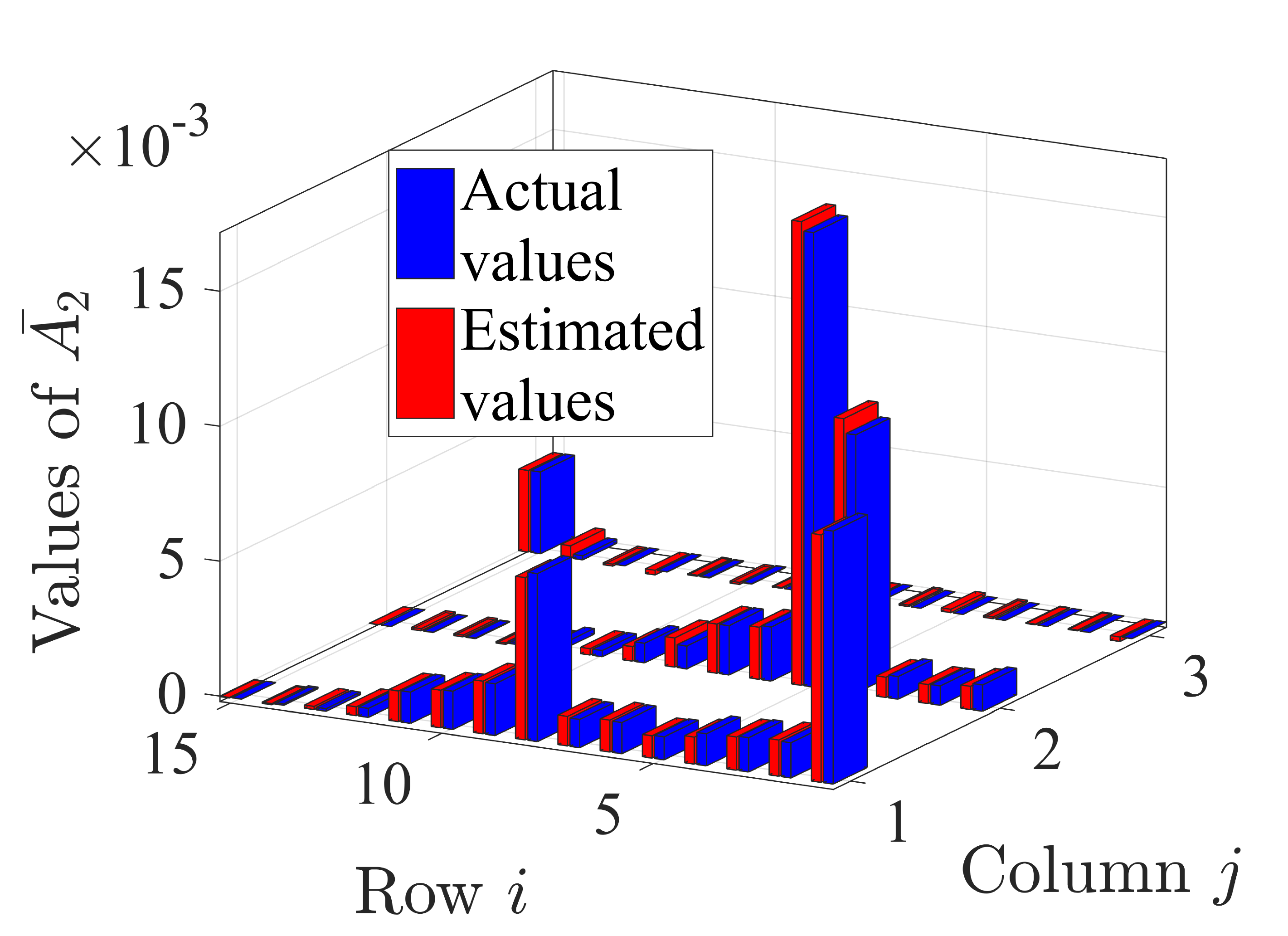}
\end{minipage}%
}%
\centering
\caption{The actual and estimated values of the dynamic components.}
\label{dynamic components} 
\end{figure}

State variables ${\bm{\delta }}$ and ${\bm{\omega }}$ are obtained from the emulated PMU data with a sampling rate of 50 Hz and a window size of 300 s.
 For example, Fig.~\ref{stochastics} presents the time-domain trajectory of G1's rotor speed before and after adding the small perturbation introduced in {\bf{Step 2}} by (21). It can be seen that the perturbation needed for estimating matrices is small and will not have an obvious impact on system performance. Following the procedure of the WAMS-based method of estimating matrices, the dynamic components are estimated and compared with their true values, as shown in Fig.~\ref{dynamic components}.  
All matrices can be estimated with fairly good accuracy. Particularly, the entries with large values can always be well estimated, though some discrepancies may exist in the entries with smaller values, which, nevertheless, have little impact on the performance of the designed controllers as shown in the subsequent section. {\color{black}Besides, the impacts of PMU noise and missing PMUs on the matrix estimation using (\ref{eq:estA_c}) have been discussed in \cite{ref:sheng2019online}, which shows that the closed-loop system state matrix $A_c$ or its sub-matrix can still be well estimated. Readers are referred to \cite{ref:sheng2019online} for more details.  
The subsequent discussions will focus on the effectiveness and the adpaptiveness of the proposed WAMS-Based WADC. }

\subsection{Validation For the WAMS-Based WADC}

\begin{figure} [htbp]
\centering
\subfigure[Eigenvalues of the estimated system state matrix]{ \label{Estimated eigenvalues} 
\begin{minipage}[t]{0.5\linewidth}
\centering
\includegraphics[width=1.8in]{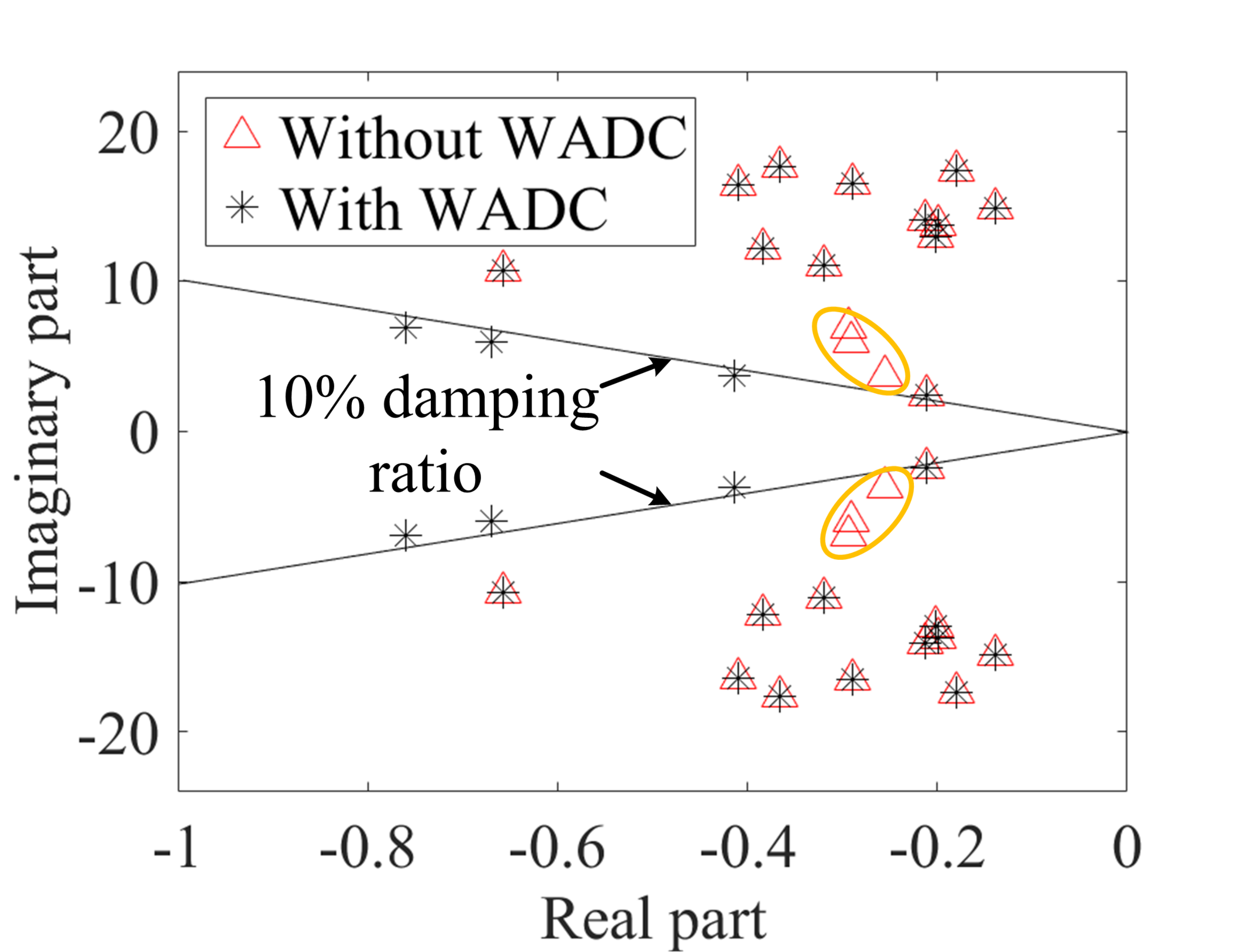}
\end{minipage}%
}%
\subfigure[Eigenvalues of the actual system state matrix]{ \label{Actual eigenvalues} 
\begin{minipage}[t]{0.5\linewidth}
\centering
\includegraphics[width=1.8in]{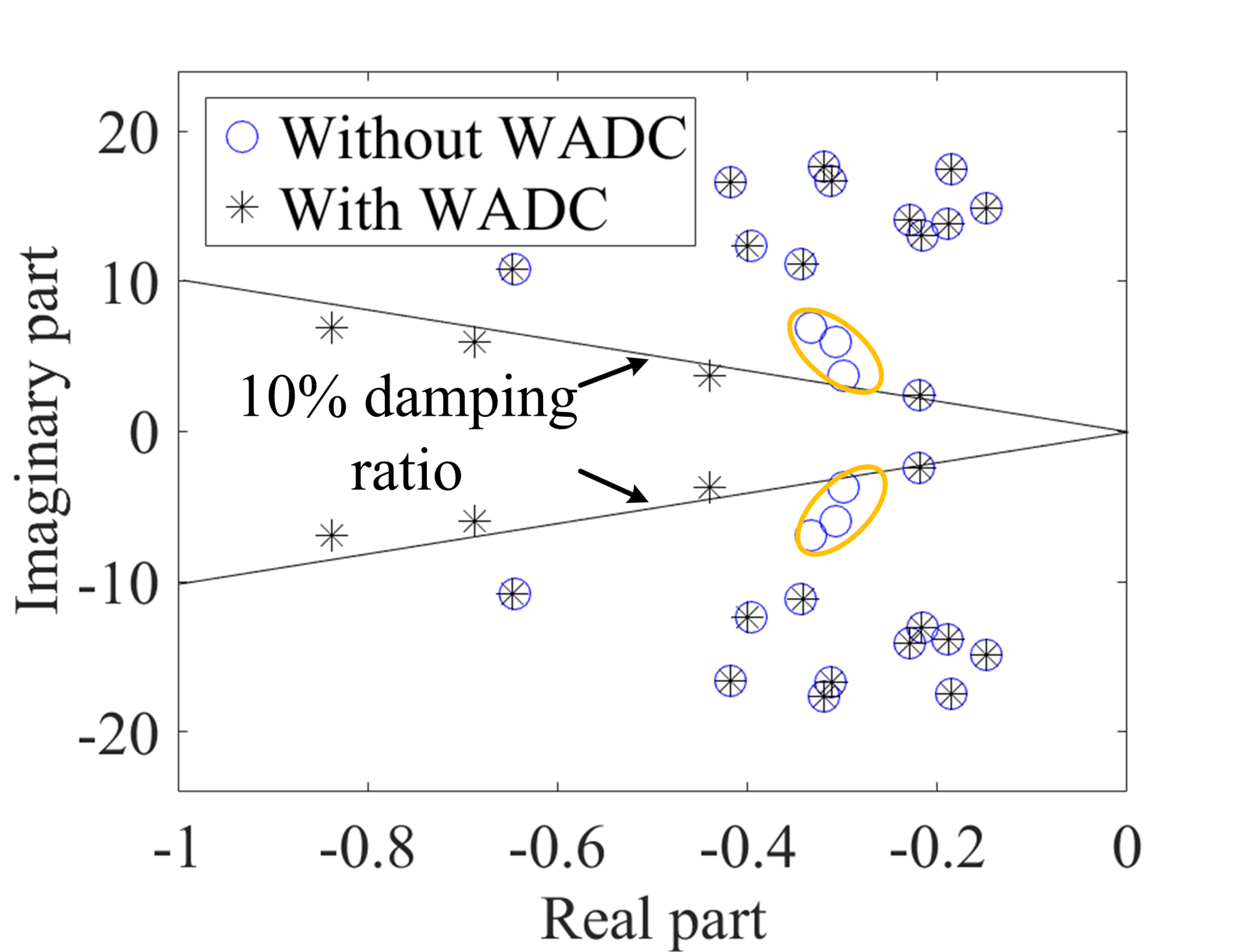}
\end{minipage}%
}%
\centering
\caption{Comparison of eigenvalues with and without WADC.}
\label{eigenvalues} 
\end{figure}

The eigenvalues of the open-loop system state matrix are denoted in blue in Fig. \ref{Actual eigenvalues}, where all inter-area modes (those in the yellow circle) are poorly damped. In order to increase the damping ratios of the inter-area oscillations to 10\%, above which the modes are considered to be well-damped \cite{ref:rogers2012power}, the WADC method introduced in section IV is applied. Specifically, the entries of the weighting matrix ${{{W}}_{{Q}}}$ corresponding to three inter-area modes are adjusted until the design objective is achieved and the other entries are set to 0. In this case, the entries corresponding to three inter-area modes are set to 2, 2.7 and 65, respectively. Besides, an identity matrix is used for the weighting matrix ${{{W}}_{{R}}}$, which assumes that all VSCs have the same modulation capacities.

Fig.~\ref{eigenvalues} presents the comparison of eigenvalues with and without applying the WADC method. The results based on the estimated matrices is shown in Fig.~\ref{Estimated eigenvalues}. It can be seen that the eigenvalues corresponding to the three critical inter-area modes all move to the left, making the damping ratios larger than 10\%. Moreover, the WADC designed based on the estimation is still effective in the true system. As shown in Fig.~\ref{Actual eigenvalues}, the damping ratios of all the critical inter-area modes of the true system have been increased above 10\% by the designed WADC based on estimated matrices, {indicating that multiple inter-area modes are damped simultaneously. In the mean time,} the rest of modes are unaffected, demonstrating the decoupling between different modes by the proposed method.
{\color{black}
In addition, the proposed WADC method took only 0.25s on a regular laptop (Core i7-7500U @ 2.70GHz, 16 GB RAM) to estimate the matrices and to design MLQR, indicating negligible computational time and good feasibility in an online environment.}

\subsection{Dynamic response to system disturbances}
To validate the effectiveness of the WAMS-based WADC method under different system disturbances, we test the performance of the proposed method in two situations---under the variation of load and generation and under a line fault. 
\subsubsection{Load and generation variation}
\begin{figure} [htbp]
\centering
\subfigure[Rotor angle of G5 relative to G13]{ \label{Rotor angle load variation} 
\begin{minipage}[t]{0.5\linewidth}
\centering
\includegraphics[width=1.8in]{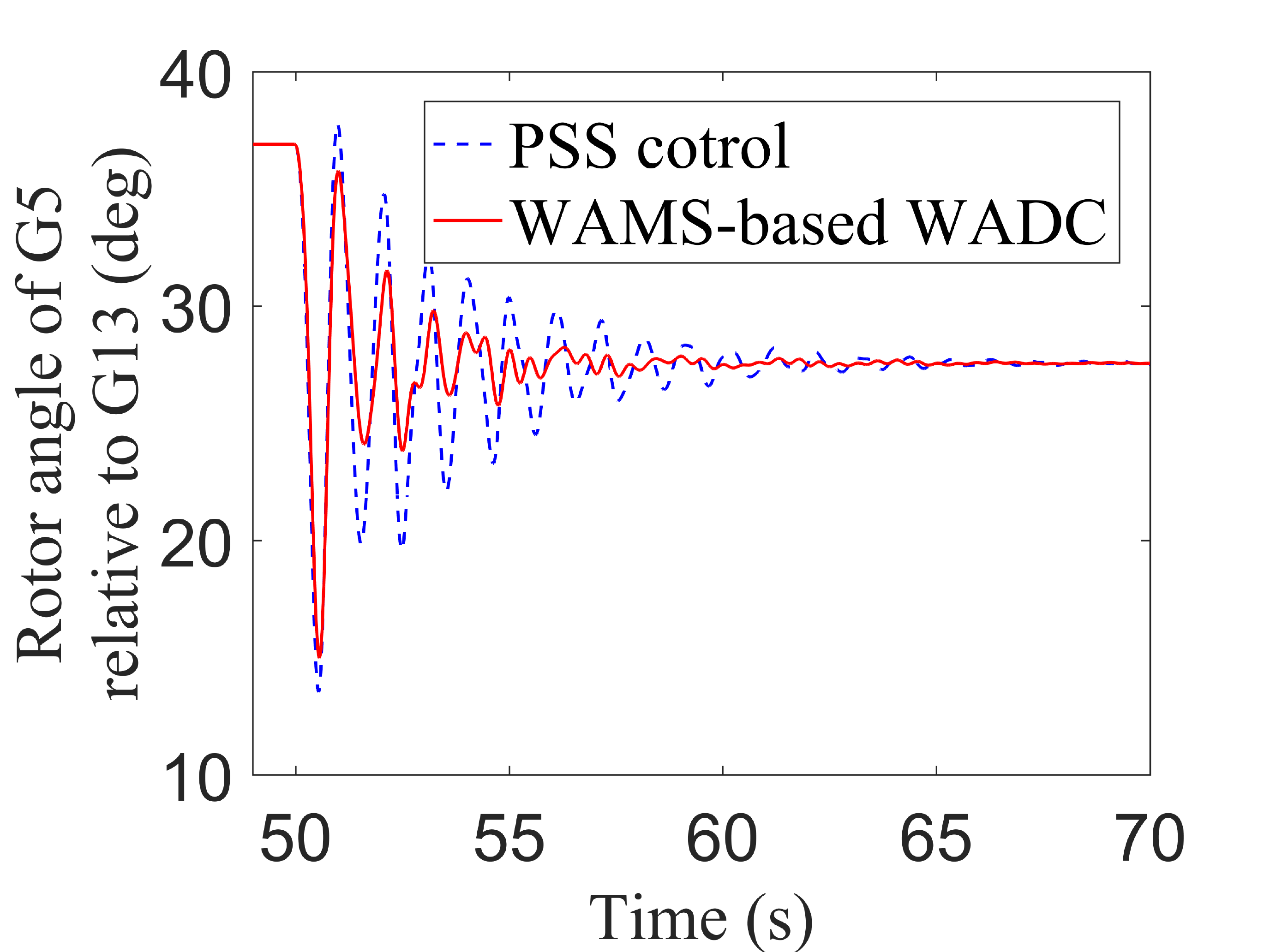}
\end{minipage}%
}%
\subfigure[Active power of VSCs]{ \label{Pvsc load variation} 
\begin{minipage}[t]{0.5\linewidth}
\centering
\includegraphics[width=1.8in]{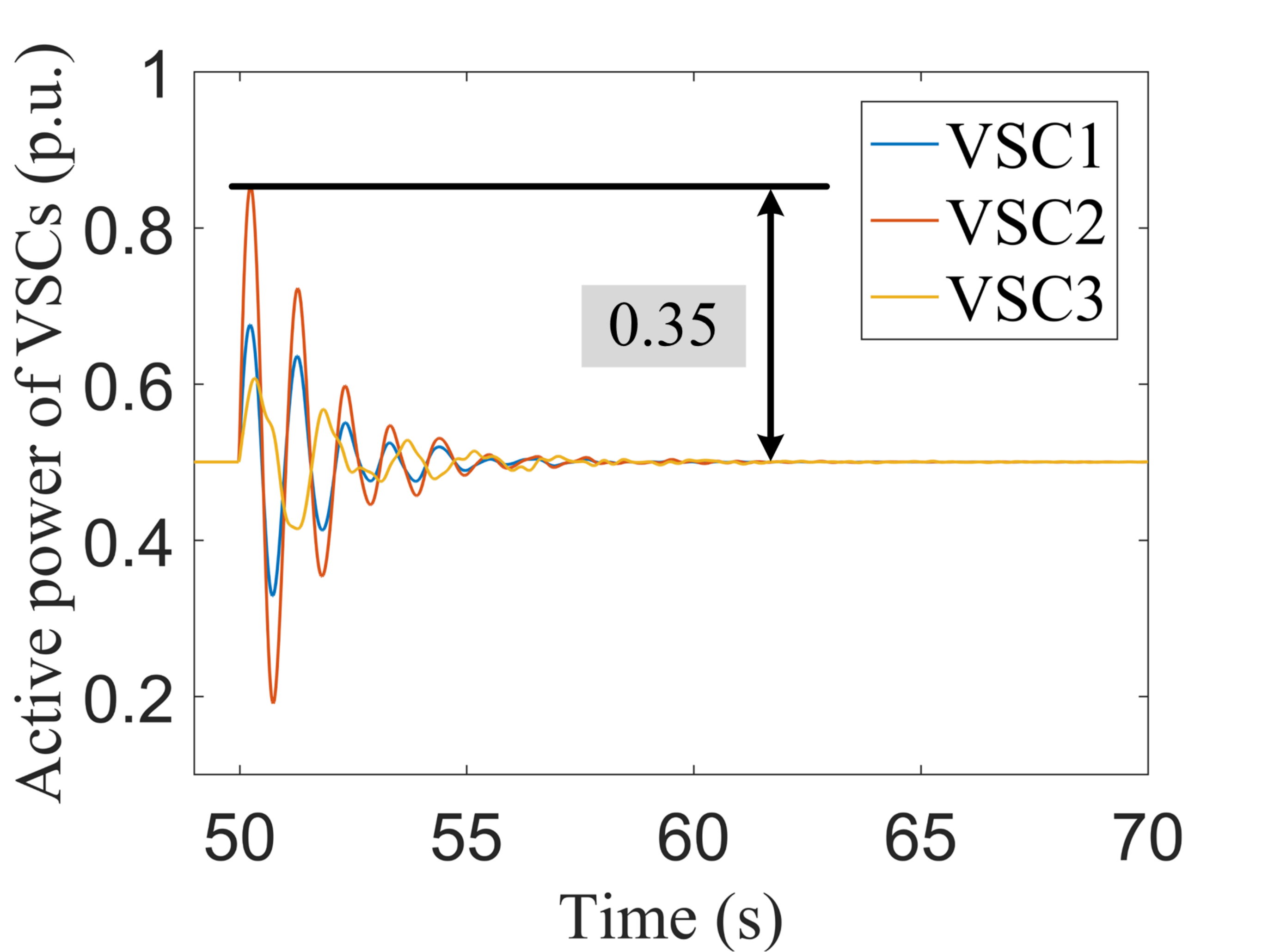}
\end{minipage}%
}%
\centering
\caption{The system performance under a generation and load variation.}
\label{load variation} 
\end{figure}

In the first case, there is a sudden 10\% generation decrease from G8-G10 and 10\% load decrease from bus 17 at 50.0 s. The dynamic response of the rotor angle difference between G5 and G13 by applying the WAMS-based WADC method is compared to that with PSS control, as shown in Fig.~\ref{Rotor angle load variation}. It can be seen that the WAMS-based WADC method seems to be more effective than PSS in damping the inter-area oscillation modes. Fig.~\ref{Pvsc load variation} presents the output active power of VSCs, showing the support of VSCs to damp the oscillations.

\subsubsection{Line fault}
\begin{figure} [htbp]
\centering
\subfigure[The rotor angle of G5 relative to G13]{ \label{Rotor angle fault} 
\begin{minipage}[t]{0.5\linewidth}
\centering
\includegraphics[width=1.8in]{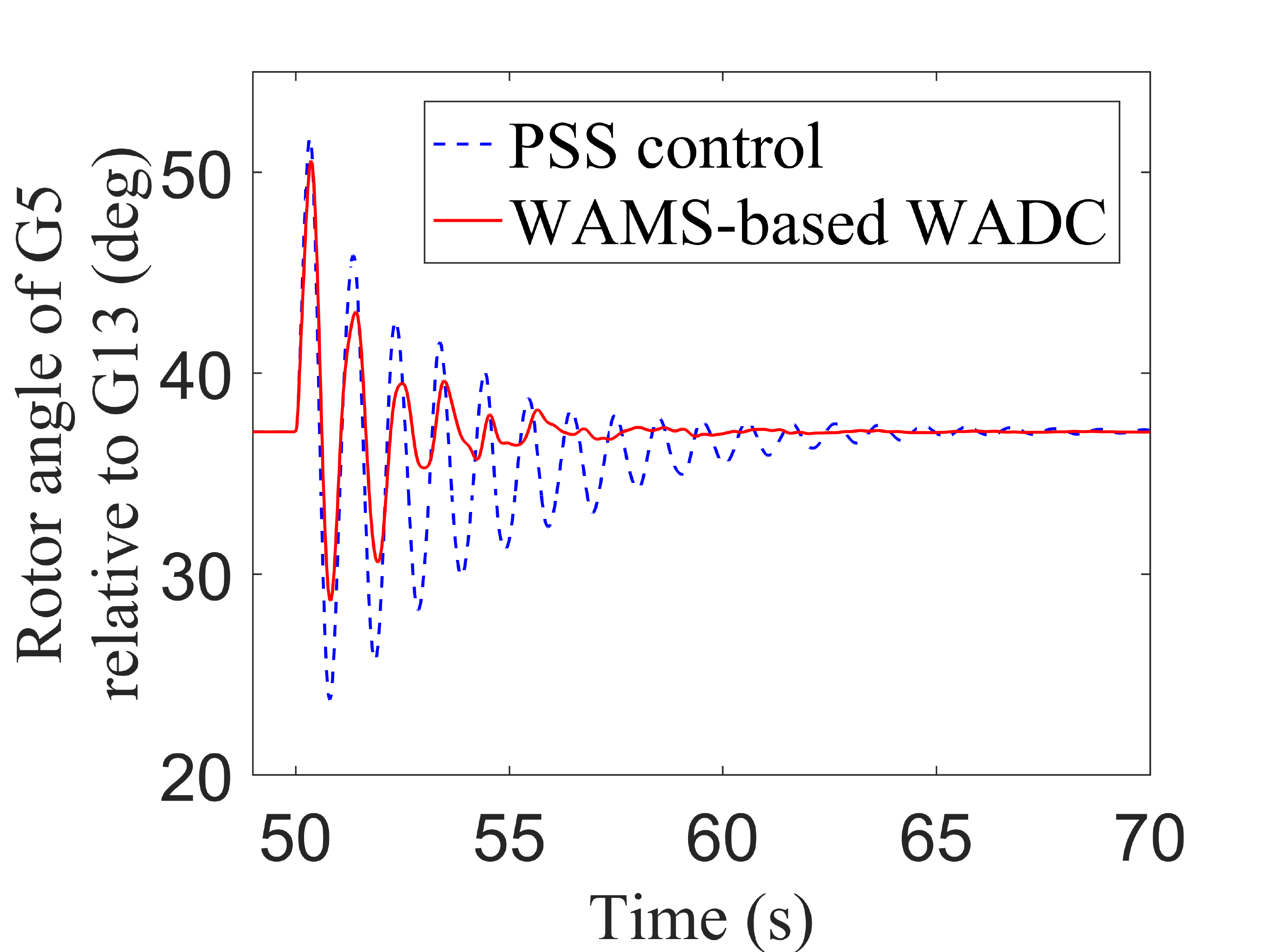}
\end{minipage}%
}%
\subfigure[The active power of VSCs]{ \label{Pvsc fault} 
\begin{minipage}[t]{0.5\linewidth}
\centering
\includegraphics[width=1.8in]{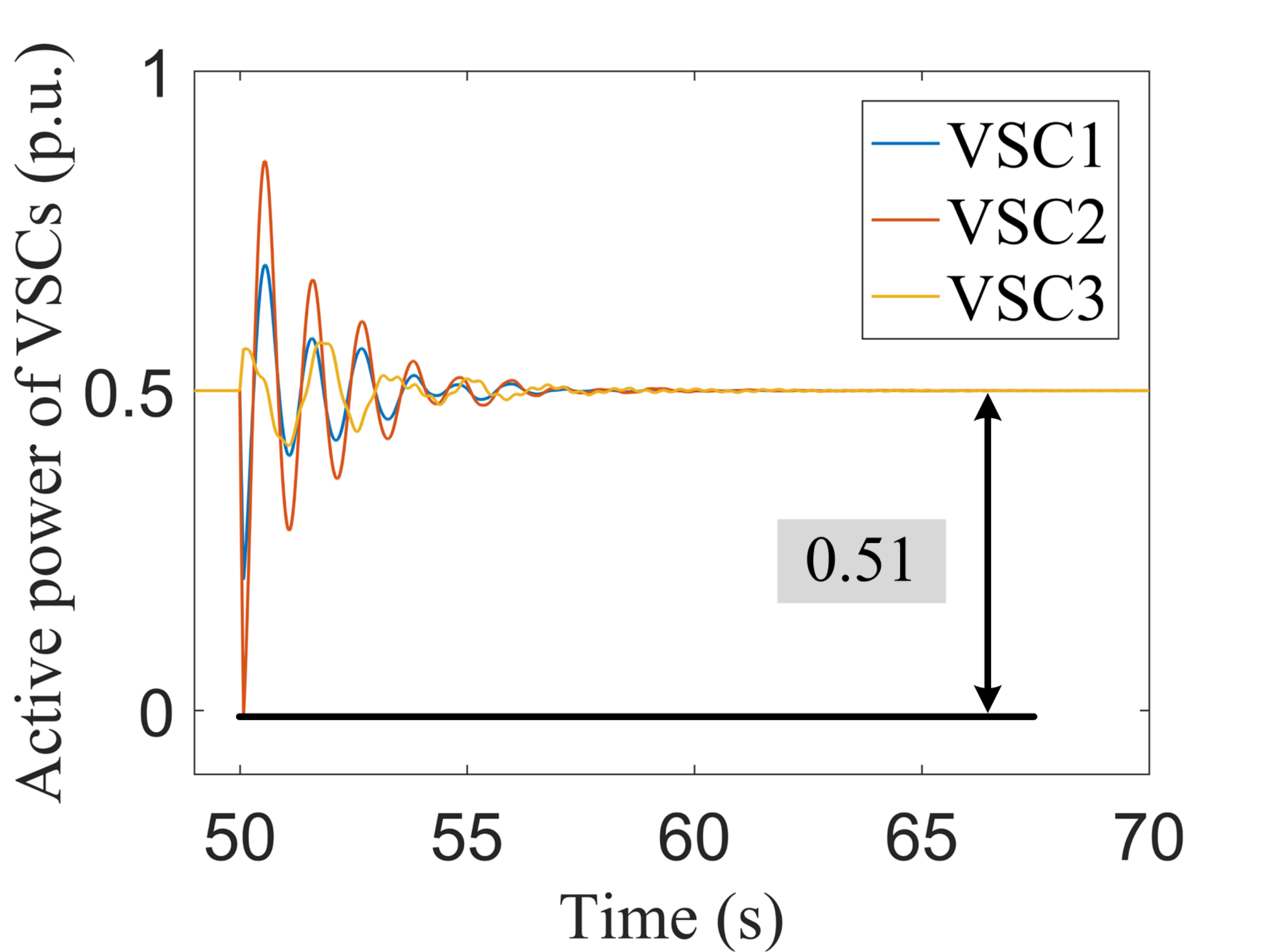}
\end{minipage}%
}%
\centering
\caption{The system performance when there is a line-to-ground fault at bus 54.}
\label{fault} 
\end{figure}

In the second case, a three-phase line-to-ground fault is applied to bus 54 at 50.0 s and is cleared after 0.0833s. Similarly, the dynamic response of the relative angle between G5 and G13 is presented in Fig.~\ref{Rotor angle fault}. Obviously, a better damping performance is achieved using the proposed WAMS-based WADC method compared to the conventional PSS control. In addition, comparing the results in Fig.~\ref{Pvsc fault} with those in Fig.~\ref{Pvsc load variation}, we note that the maximum modulated power of VSC2 is increased from 0.35 p.u. to 0.51 p.u. when providing the damping support given that the line fault may be a more severe disturbance compared to the power variation of load and generation. 

{\color{black}It should be noted that the communication delay in practical applications may deteriorate the damping performance, which, however, can be alleviated by appropriate compensation approaches (e.g., \cite{ref:chaudhuri2009new}).} 

\newcommand{\tabincell}[2]{\begin{tabular}{@{}#1@{}}#2\end{tabular}}  

\begin{table*} 
\captionsetup{justification=centering, labelsep=newline}
\centering
 \caption{An electromechanical mode (Mode 2) under different control strategies and working conditions \color{black}when VSCs are located at buses 54, 20, 42\color{black}}
\label{tab1}
\resizebox{\textwidth}{!}{
\begin{tabular}{c c c c c c c}
\hline
\multirow{2}{*}{Working condition} & \multicolumn{2}{c}{Open-loop
(without damping control)
} & \multicolumn{2}{c}{Closed-loop
(the model-based WADC)
} & \multicolumn{2}{c}{Closed-loop
(the WAMS-based WADC)
}\\
\cline{2-7}
& Frequency (Hz) & Damping ratio (\%) & Frequency (Hz) & Damping ratio (\%) & Frequency (Hz) & Damping ratio (\%) \\ 
\hline
Base case  & 0.952 & 5.121 & 0.943 & 11.143 & 0.943 & 11.470\\
\tabincell{c} {Line outage at bus 60-61} & 0.735 &	4.231 &	0.724 &	8.872 &	0.720 &	11.900\\
\tabincell{c} {Line outage at bus 53-54} & 0.757 &	4.107 &	0.746 &	8.800 &	0.743 &	11.643\\
\textcolor{black}{\tabincell{c} {Generation and load variation\\ followed by a line outage 53-54}} & \textcolor{black}{0.727} & \textcolor{black}{3.780} &	\textcolor{black}{0.717} &	\textcolor{black}{7.989} &	\textcolor{black}{0.715} &	\textcolor{black}{11.209}\\
\hline
\end{tabular}}
\end{table*}

\subsection{Adaptiveness to different working conditions}
In contrast to the model-based WADC, which {\color{black}may not} consider the change of working conditions, a significant advantage of the proposed WAMS-based WADC method is that the damping coefficients can be adjusted as the working condition varies. {\color{black}To show this, the adaptiveness of the proposed method  to the individual line outage and a combined case where both line outage and load variations occur will be demonstrated. Specifically, three working conditions except the base case are considered: the condition after a line outage at bus 60-61; the condition after a line outage at bus 53-54;  the condition after a sudden 10\% load increase from bus 17 and 10\% generation increase from G8-G10, followed by a line outage at bus 53-54 1s afterward.} 

\begin{figure} [htbp]
\centering
\subfigure[The condition after the line outage at bus 60-61]{ \label{mode2-bus6061} 
\begin{minipage}[t]{0.5\linewidth}
\centering
\includegraphics[width=1.8in]{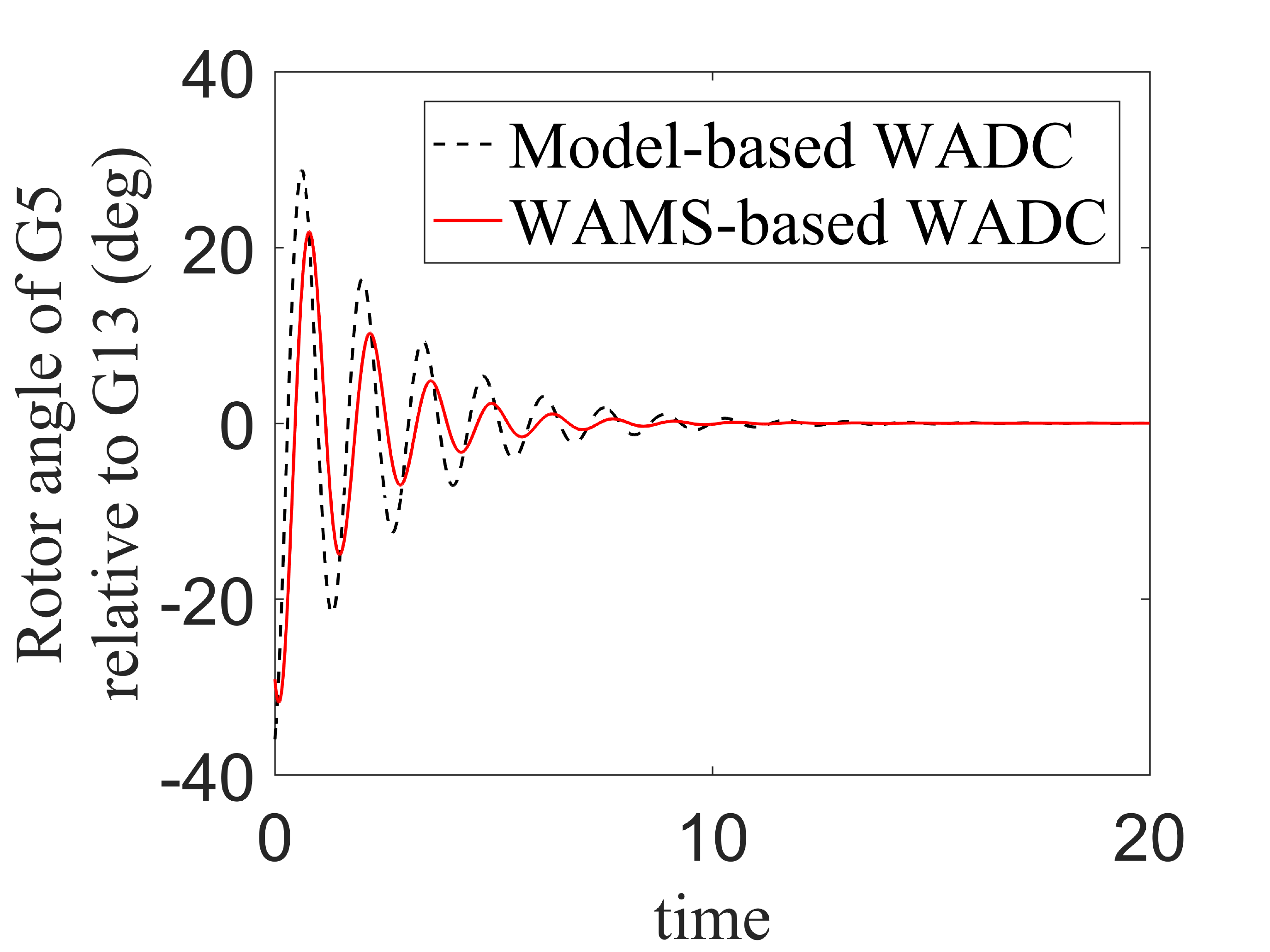}
\end{minipage}%
}%
\subfigure[The condition after the line outage at bus 53-54]{ \label{mode2-bus5354} 
\begin{minipage}[t]{0.5\linewidth}
\centering
\includegraphics[width=1.8in]{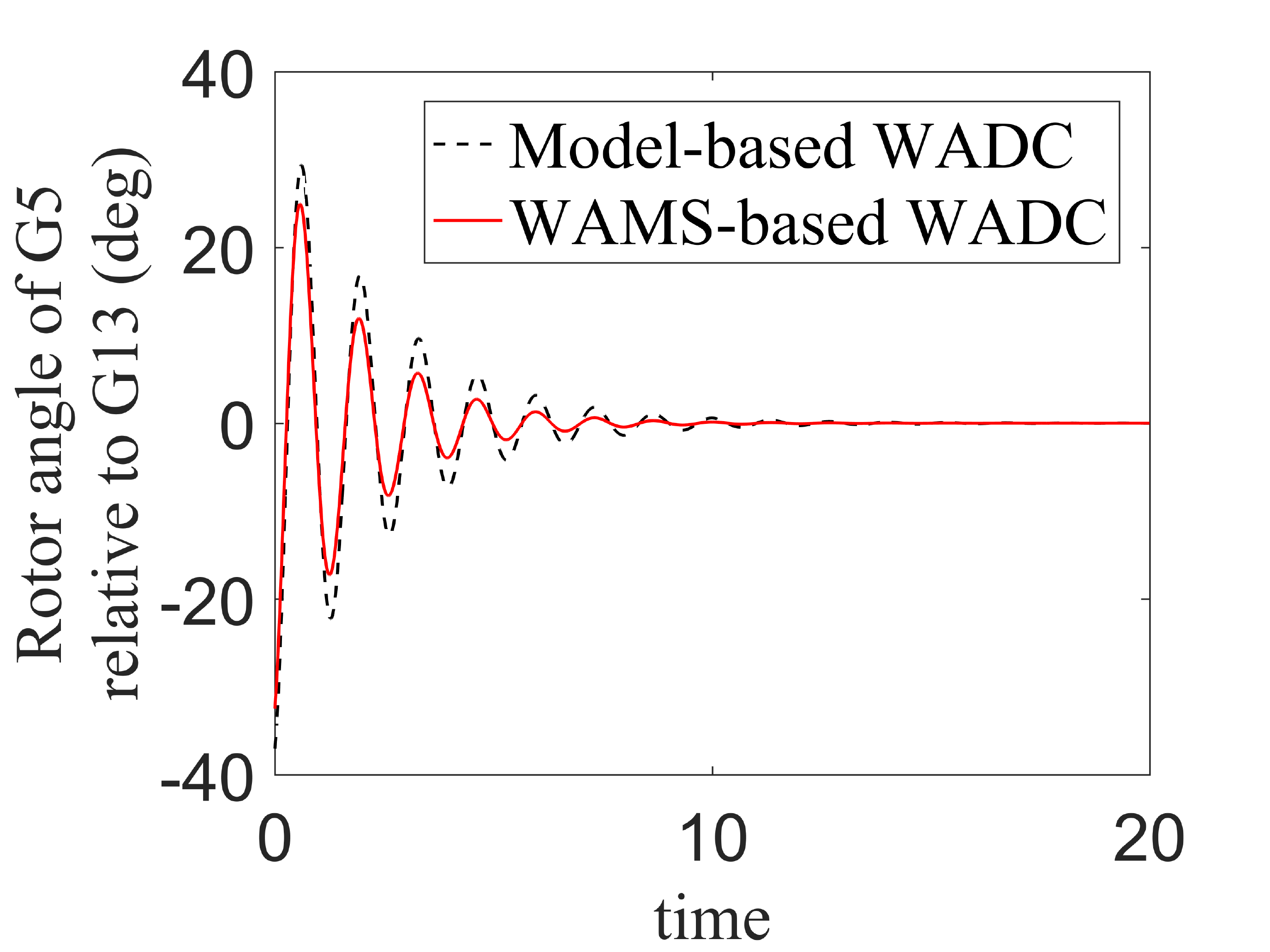}
\end{minipage}%
}%

\subfigure[{\color{black}The condition after generation and load variations followed by a line outage at bus 53-54}]{ \label{mode2-generation-load-bus5354} 
\begin{minipage}[t]{0.5\linewidth}
\centering
\includegraphics[width=1.8in]{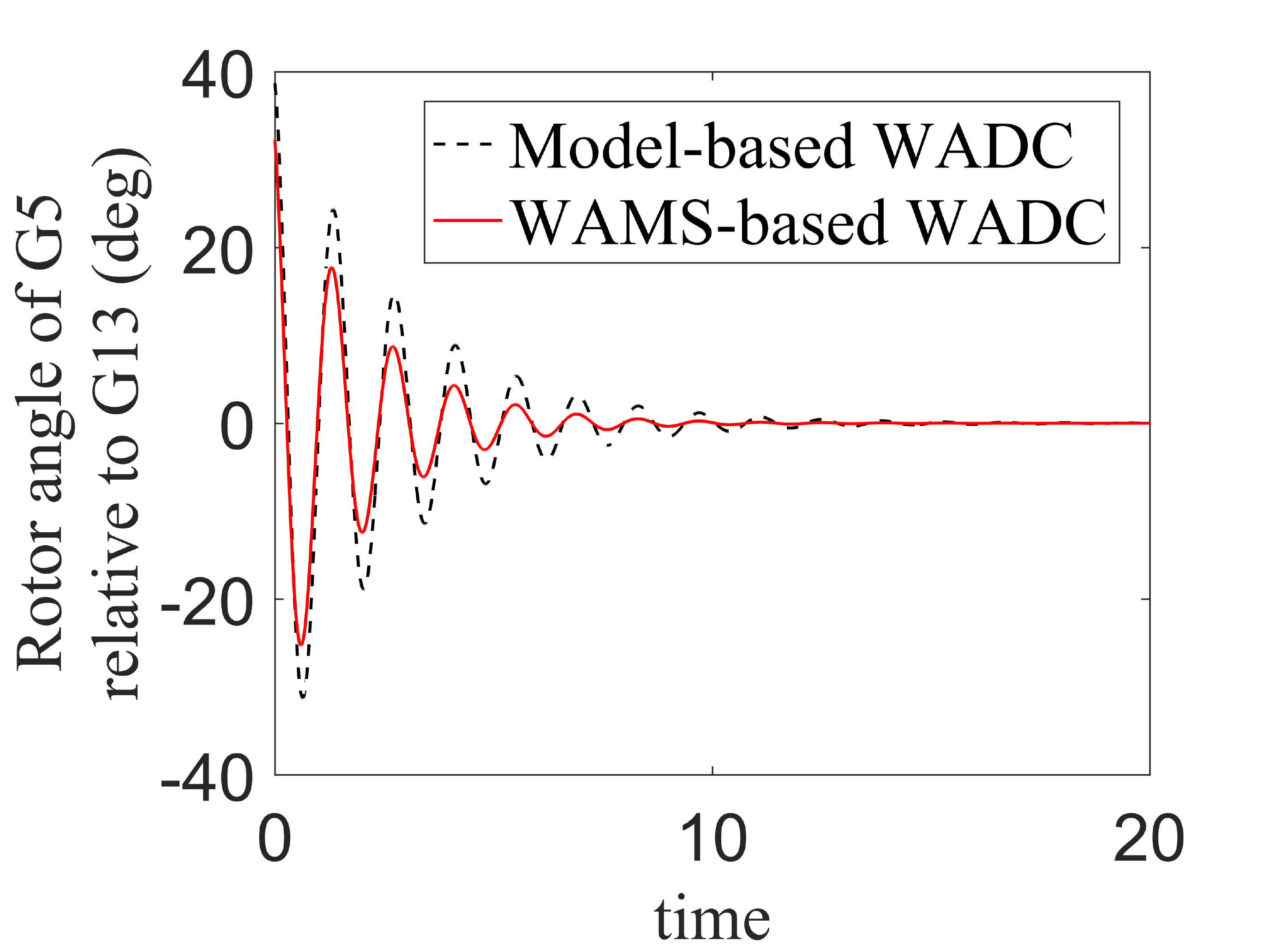}
\end{minipage}%
}%
\centering
\caption{The system response to Mode 2 in different working conditions with different control methods.}
\label{Adapativeness} 
\end{figure}

The system performance without any control is compared to that with the model-based WADC using MLQR and that with the WAMS-based WADC using MLQR. We assume that \color{black}the topology changes are undetected such that \color{black}the model-based WADC designed for the base case 
remains unchanged. Table~\ref{tab1} presents the frequency and the damping ratio of an inter-area oscillation (Mode 2) under different control strategies and in various working conditions. It can be seen that both the model-based WADC and the WAMS-based WADC can increase the damping ratio to above 10\% in the base case. {\color{black}However, Mode 2 changes from well-damped to poorly-damped when any of the aforementioned working conditions happens, whereas the model-based WADC fails to increase the damping ratio of Mode 2 to the requested 10\%. The model-based WADC becomes ineffective when the working condition changes as the controller damping gain is not updated due to the undetected topology changes or load variations.}
In contrast, the WAMS-based WADC can always ensure that the minimum requested damping ratio is achieved   {\color{black} when the working condition and/or the characteristics of electromechanical modes change.} 
In addition, Fig.~\ref{Adapativeness} shows comparisons of  the time domain response {\color{black}when Mode 2 (0.952Hz) is excited in different working conditions with different control strategies}.  Obviously, we can see that the WAMS-based WADC can always achieve better damping performance than the model-based WADC. Note that the difference is not huge because the targeted damping ratio in the WAMS-based WADC is 10\%. A higher damping ratio can be achieved by setting a higher threshold.

\color{black}
\subsection{Validation for different VSCs' locations} \label{section_diffVSCs}

\begin{figure} [htbp]
\captionsetup{labelfont={color=black},font={color=black}}
\centering
\subfigure[Eigenvalues of the estimated system state matrix]{ \label{New Estimated eigenvalues} 
\begin{minipage}[t]{0.5\linewidth}
\centering
\includegraphics[width=1.8in]{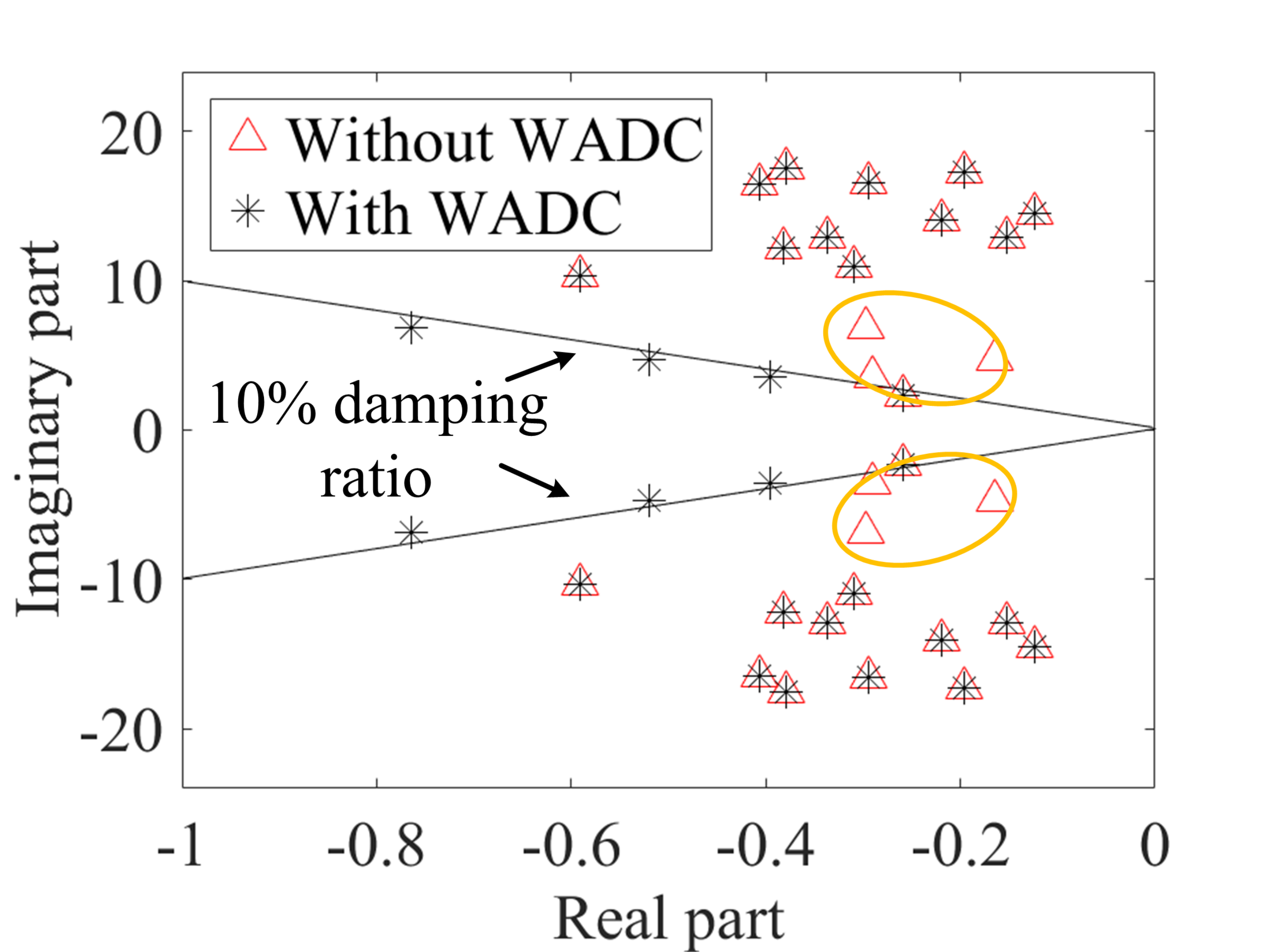}
\end{minipage}%
}%
\subfigure[Eigenvalues of the actual system state matrix]{ \label{New Actual eigenvalues} 
\begin{minipage}[t]{0.5\linewidth}
\centering
\includegraphics[width=1.8in]{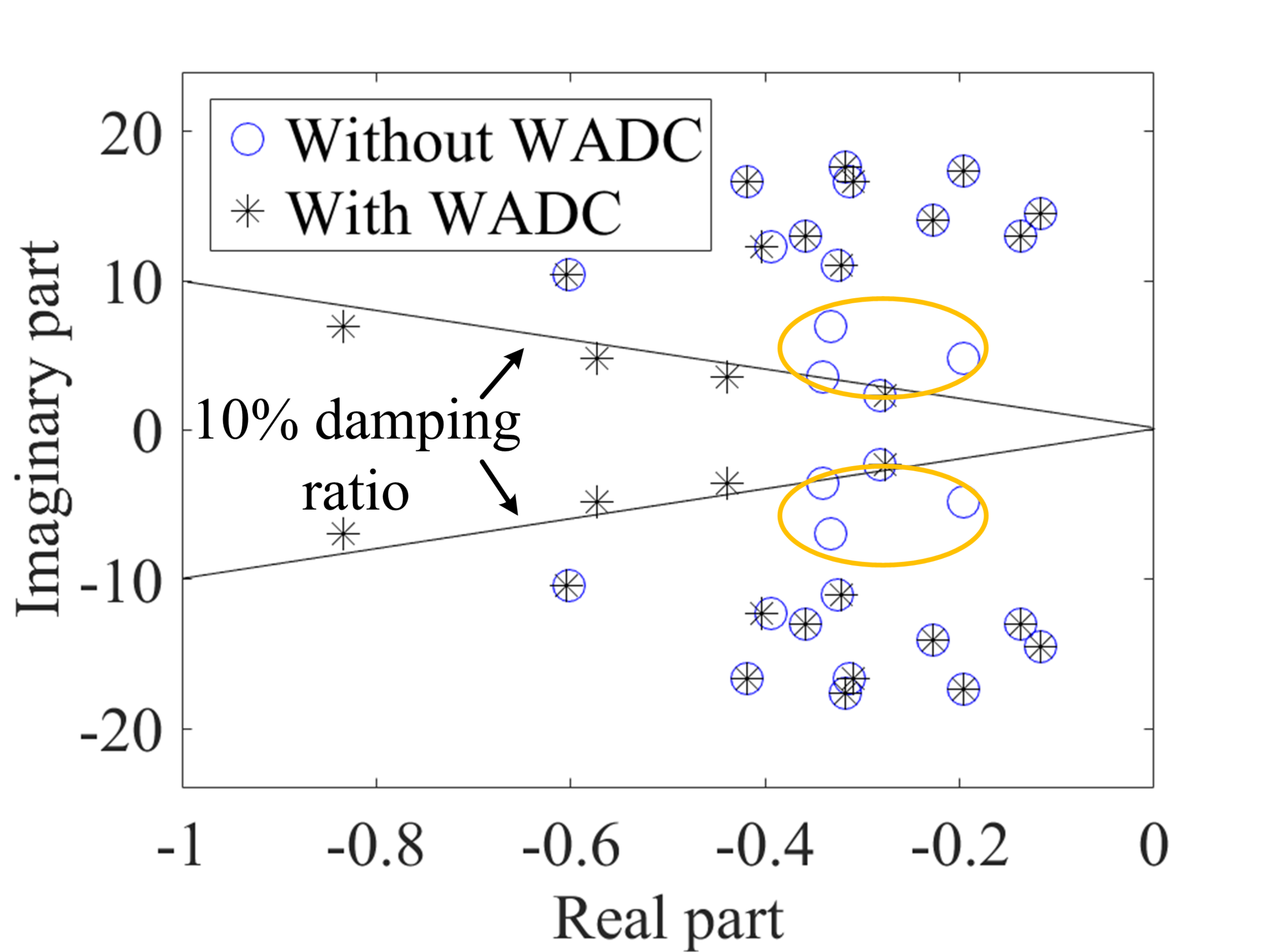}
\end{minipage}%
}%
\centering
\caption{Comparison of eigenvalues with and without WADC when VSCs' locations change (at buses 56, 20, 42).}
\label{new eigenvalues} 
\end{figure}


\begin{table*} 
\captionsetup{labelfont={color=black},font={color=black}, justification=centering, labelsep=newline}
\centering
\caption{An electromechanical mode (Mode 2) under different control strategies and working conditions when VSCs are located at buses 56, 20, 42.}
 \label{tab2}
 \textcolor{black}{
\resizebox{\textwidth}{!}{
\begin{tabular}{c c c c c c c}
\hline
\multirow{2}{*}{Working condition} & \multicolumn{2}{c}{Open-loop
(without damping control)
} & \multicolumn{2}{c}{Closed-loop
(the model-based WADC)
} & \multicolumn{2}{c}{Closed-loop
(the WAMS-based WADC)
}\\
\cline{2-7}
& Frequency (Hz) & Damping ratio (\%) & Frequency (Hz) & Damping ratio (\%) & Frequency (Hz) & Damping ratio (\%) \\ 
\hline
Base case  & 0.952 & 5.119 & 0.943 & 11.216 & 0.942 & 11.539\\
\tabincell{c} {Line outage at bus 53-54} & 0.759 &	4.113 &	0.749 &	8.400 &	0.745 &	11.876\\
{\tabincell{c} {Generation and load variation\\ followed by a line outage 53-54}} & 0.729 & 3.796 &	0.720 &	7.656 &	0.714 &	12.412\\
\hline
\end{tabular}}}
\end{table*}

In practice, VSCs may be located electrically far from the generations. As a result, the effectiveness of the proposed strategy by using VSCs at different locations is tested in this section. Generally speaking, the selection of the locations of VSCs can follow the controllability indexes.  
Interested readers can check more details in \cite{ref:latorre2011improvement, ref:trinh2016analytical}. The controllability analysis shows that VSCs added to bus 56 and 20 have a relative large controllability indexes associated with Mode 1 (0.594Hz) and Mode 2 (0.952Hz). In comparison, the VSC added to bus 42 is more effective in Mode 1 and Mode 3 (1.108Hz). Therefore, a system with three VSCs placed at bus 56, 20 and 42 is considered, which has VSCs both far from and close to generators. 
More rigorous analysis on how to select the feedback signals and optimize the locations of VSCs for an effective control of multiple modes requires future effort.

Modal analysis shows that there are still three critical oscillation modes when damping control is not provided by VSCs. After applying the proposed WADC method, the damping ratios of the three modes can be successfully increased to above 10\%, as presented in Fig.~\ref{new eigenvalues}. Additionally, the adaptiveness of the WAMS-based WADC strategy is tested in three working conditions: the base working condition, the condition after a line outage at bus 53-54 as well as the condition after a generation and load variation followed by the line outage at bus 53-54, which is the same as the one applied in Section IV.D. The results summarized in Table.~\ref{tab2} validate that the proposed WAMS-based WADC control can always maintain the required damping ratio, which {\color{black}may not} be achieved by the model-based WADC in different working conditions.

\color{black}

\section{Conclusion and Future Work}
This paper has proposed a novel model-free WADC method to damp multiple inter-area oscillations using VSCs in various operation conditions.  
The method can be divided into two steps: the WAMS-based model-free {estimation of the actual system state matrix $A$ and the input matrix $B$, which is followed by} the MLQR-based WADC method using the estimated matrices. The proposed method, being independent of model parameters and network topology, can adjust the control signals of VSCs to provide sufficient damping as the system evolves.
Numerical studies in the modified IEEE 68-bus system show that the inter-area oscillations can always be well damped by the proposed method regardless of the change of working conditions and network topology. In the near future, robust control methods with respect to time delay will be studied. {\color{black} More detailed VSC dynamics and optimal placement of VSCs to provide damping will also be considered.}  

\appendices
\small
\section{} \label{Aderivation}
Firstly, the matrix of derivatives in (\ref{eq10}) are expressed by
\begin{equation}
\left[ \begin{array}{l}
\frac{{\partial {{\bm{P}}_{{E}}}}}{{\partial {{\delta }}}}{\kern 1pt} {\kern 1pt} {\kern 1pt} {\kern 1pt} {\kern 1pt} {\kern 1pt} {\kern 1pt} {\kern 1pt} \frac{{\partial {{\bm{P}}_{{E}}}}}{{\partial {\bm{\theta }}}}{\kern 1pt} {\kern 1pt} {\kern 1pt} {\kern 1pt} {\kern 1pt} {\kern 1pt} {\kern 1pt} {\kern 1pt} \frac{{\partial {{\bm{P}}_{{E}}}}}{{\partial {\bm{V}}}}\\
\frac{{\partial {{\bm{P}}_{\bm{v}}}}}{{\partial {{\delta }}}}{\kern 1pt} {\kern 1pt} {\kern 1pt} {\kern 1pt} {\kern 1pt} {\kern 1pt} {\kern 1pt} {\kern 1pt} {\kern 1pt} \frac{{\partial {{\bm{P}}_{\bm{v}}}}}{{\partial {\bm{\theta }}}}{\kern 1pt} {\kern 1pt} {\kern 1pt} {\kern 1pt} {\kern 1pt} {\kern 1pt} {\kern 1pt} {\kern 1pt} \frac{{\partial {{\bm{P}}_{\bm{v}}}}}{{\partial {\bm{V}}}}\\
\frac{{\partial {{\bm{Q}}_{\bm{v}}}}}{{\partial {{\delta }}}}{\kern 1pt} {\kern 1pt} {\kern 1pt} {\kern 1pt} {\kern 1pt} {\kern 1pt} \frac{{\partial {{\bm{Q}}_{\bm{v}}}}}{{\partial {\bm{\theta }}}}{\kern 1pt} {\kern 1pt} {\kern 1pt} {\kern 1pt} {\kern 1pt} {\kern 1pt} \frac{{\partial {{\bm{Q}}_{\bm{v}}}}}{{\partial {\bm{V}}}}
\end{array} \right] = \left[ \begin{array}{l}
{{{A}}_{{{11}}}}{\kern 1pt} {\kern 1pt} {\kern 1pt} {\kern 1pt} {\kern 1pt} {\kern 1pt} {\kern 1pt} {\kern 1pt} {\kern 1pt} {\kern 1pt} {{{A}}_{{{12}}}}{\kern 1pt} {\kern 1pt} {\kern 1pt} {\kern 1pt} {\kern 1pt} {\kern 1pt} {\kern 1pt} {\kern 1pt} {{{A}}_{{{13}}}}\\
{{{A}}_{{{21}}}}{\kern 1pt} {\kern 1pt} {\kern 1pt} {\kern 1pt} {\kern 1pt} {\kern 1pt} {\kern 1pt} {\kern 1pt} {\kern 1pt} {{{A}}_{{{22}}}}{\kern 1pt} {\kern 1pt} {\kern 1pt} {\kern 1pt} {\kern 1pt} {\kern 1pt} {\kern 1pt} {\kern 1pt} {{{A}}_{{{23}}}}\\
{{{A}}_{{{31}}}}{\kern 1pt} {\kern 1pt} {\kern 1pt} {\kern 1pt} {\kern 1pt} {\kern 1pt} {\kern 1pt} {\kern 1pt} {\kern 1pt} {{{A}}_{{{32}}}}{\kern 1pt} {\kern 1pt} {\kern 1pt} {\kern 1pt} {\kern 1pt} {\kern 1pt} {\kern 1pt} {\kern 1pt} {{{A}}_{{{33}}}}
\end{array} \right] \label{eq32}
\end{equation}

Based on the second and third rows of (\ref{eq10}), ${{\Delta \bm{\theta} }}$ and ${{\Delta \bm{V}}}$ can be calculated by
\begin{equation}
\begin{aligned}
{{\Delta \bm{\theta} }} &= {F_1}{{{A}}_{{{23}}}}^{ - 1}\left( {{{\Delta }}{{\bm{P}}_{\bm{v}}} - {{{A}}_{{{21}}}}{{\Delta \bm{\delta} }}} \right) \\
&- {F_1}{{{A}}_{{{33}}}}^{ - 1}\left( {{{\Delta }}{{\bm{Q}}_{\bm{v}}} - {{{A}}_{{{31}}}}{{\Delta \bm{\delta} }}} \right)\\
{{\Delta \bm{V}}} &= {F_2}{{{A}}_{{{22}}}}^{ - 1}\left( {{{\Delta }}{{\bm{P}}_{\bm{v}}}{\bm{ - }}{{{A}}_{{{21}}}}{{\Delta \bm{\delta} }}} \right) \\
&- {F_2}{{{A}}_{{{32}}}}^{ - 1}\left( {{{\Delta }}{{\bm{Q}}_{\bm{v}}} - {{{A}}_{{{31}}}}{{\Delta \bm{\delta} }}} \right)
\end{aligned} \label{eq33}
\end{equation}
where \begin{equation} \nonumber
\begin{aligned} 
{F_1} &= {\left( {{{{A}}_{{{23}}}}^{ - 1}{{{A}}_{{{22}}}} - {{{A}}_{{{33}}}}^{ - 1}{{{A}}_{{{32}}}}} \right)^{ - 1}} \\
{F_2} &= {\left( {{{{A}}_{{{22}}}}^{ - 1}{{{A}}_{{{23}}}} - {{{A}}_{{{32}}}}^{ - 1}{{{A}}_{{{33}}}}} \right)^{ - 1}}
\end{aligned}    
\end{equation}

According to the first row of (\ref{eq10}) and substituting ${{\Delta \bm{\theta} }}$ and ${{\Delta \bm{V}}}$ by (\ref{eq33}), we have
\begin{equation} 
\begin{aligned}
{{\Delta }}{{\bm{P}}_{{E}}} = {{{A}}_{{{11}}}}{{\Delta \bm{\delta} }} + {{{A}}_{{{12}}}}{{\Delta \bm{\theta} }} + {{{A}}_{{{13}}}}{{\Delta \bm{V}}}\\
 = {{{A}}_{{1}}}{{\Delta \bm{\delta} }} + {{{A}}_{{2}}}{{\Delta }}{{\bm{P}}_{\bm{v}}} + {{{A}}_{{3}}}{{\Delta }}{{\bm{Q}}_{\bm{v}}}
\end{aligned}  \label{eq34}
\end{equation}
where \begin{equation} \nonumber 
\begin{aligned}
{{{A}}_{{1}}} &= {{{A}}_{{{11}}}} + {{{A}}_{{{12}}}}{F_1}\left( { - {{{A}}_{{{23}}}}^{ - 1}{{{A}}_{{{21}}}} + {{{A}}_{{{33}}}}^{ - 1}{{{A}}_{{{31}}}}} \right) \\
&+ {{{A}}_{{{13}}}}{F_2}\left( { - {{{A}}_{{{22}}}}^{ - 1}{{{A}}_{{{21}}}} + {{{A}}_{{{32}}}}^{ - 1}{{{A}}_{{{31}}}}} \right)\\
{{{A}}_{{2}}} &= {{{A}}_{{{12}}}}{F_1}{{{A}}_{{{23}}}}^{ - 1} + {{{A}}_{{{13}}}}{F_2}{{{A}}_{{{22}}}}^{ - 1} \\
{{{A}}_{{3}}} &=  - {{{A}}_{{{12}}}}{F_1}{{{A}}_{{{33}}}}^{ - 1} - {{{A}}_{{{13}}}}{F_2}{{{A}}_{{{32}}}}^{ - 1}
\end{aligned}
\end{equation}

\section{} \label{Approximation}
The estimated correlation matrix and the stationary covariance matrix are given by: 
\begin{equation} 
\begin{aligned}
{\hat{R}_{\bm{xx}}\left(\tau\right)}&=\left[ \begin{array}{cc} \hat{R}_{\bm{\delta \delta}}\left(\tau\right)& \hat{R}_{\bm{\delta \omega}}\left(\tau\right)\\ \hat{R}_{\bm{\omega \delta}}\left(\tau\right)& \hat{R}_{\bm{\omega \omega}}\left(\tau\right) \end{array} \right]\\
{\hat{C}_{\bm{xx}}}&=\left[ \begin{array}{cc} \hat{C}_{\bm{\delta \delta}}&\hat{C}_{\bm{\delta \omega}}\\\hat{C}_{\bm{\omega \delta}}&\hat{C}_{\bm{\omega \omega}} \end{array} \right]
\end{aligned}  \label{Approximation of matrix 1}
\end{equation}

Each entry of $\hat{R}_{\bm{\delta \delta}}\left(\tau\right)$ and $\hat{C}_{\bm{\delta \delta}}$ can be computed by
\begin{equation} 
\begin{aligned}
{\hat{R}_{{\delta_i}{\delta_j}}\left(\tau\right)}&=\frac{1}{N}\sum_{k=1}^{N-\tau}
{\left({\delta_{\left(k+\tau \right)i}}-\bar{\delta_i}\right)}{\left({\delta_{kj} }-\bar{\delta_j}\right)}\\
{\hat{C}_{{\delta_i}{\delta_j}}}&=\frac{1}{N}\sum_{k=1}^{N}
{\left({\delta_{ki} }-\bar{\delta_i}\right)}{\left({\delta_{kj} }-\bar{\delta_j}\right)}
\end{aligned}  \label{Approximation of matrix 2}
\end{equation}
where $\tau$ is a given lagging time, $\bar{\delta_i}$ is the mean value of $\delta_i$, and $N$ is the sample size. Likewise, $\hat{R}_{\bm{\delta \omega}}\left(\tau\right)$, $\hat{R}_{\bm{\omega\delta}}\left(\tau\right)$, $\hat{R}_{\bm{\omega \omega}}\left(\tau\right)$, $\hat{C}_{\bm{\delta \omega}}$, $\hat{C}_{\bm{\omega \delta}}$ and $\hat{C}_{\bm{\omega \omega}}$ can also be calculated.

\input{Main.bbl}

\bibliographystyle{IEEEtran}

\begin{IEEEbiography}[{\includegraphics[width=1in,height=1.25in,clip,keepaspectratio]{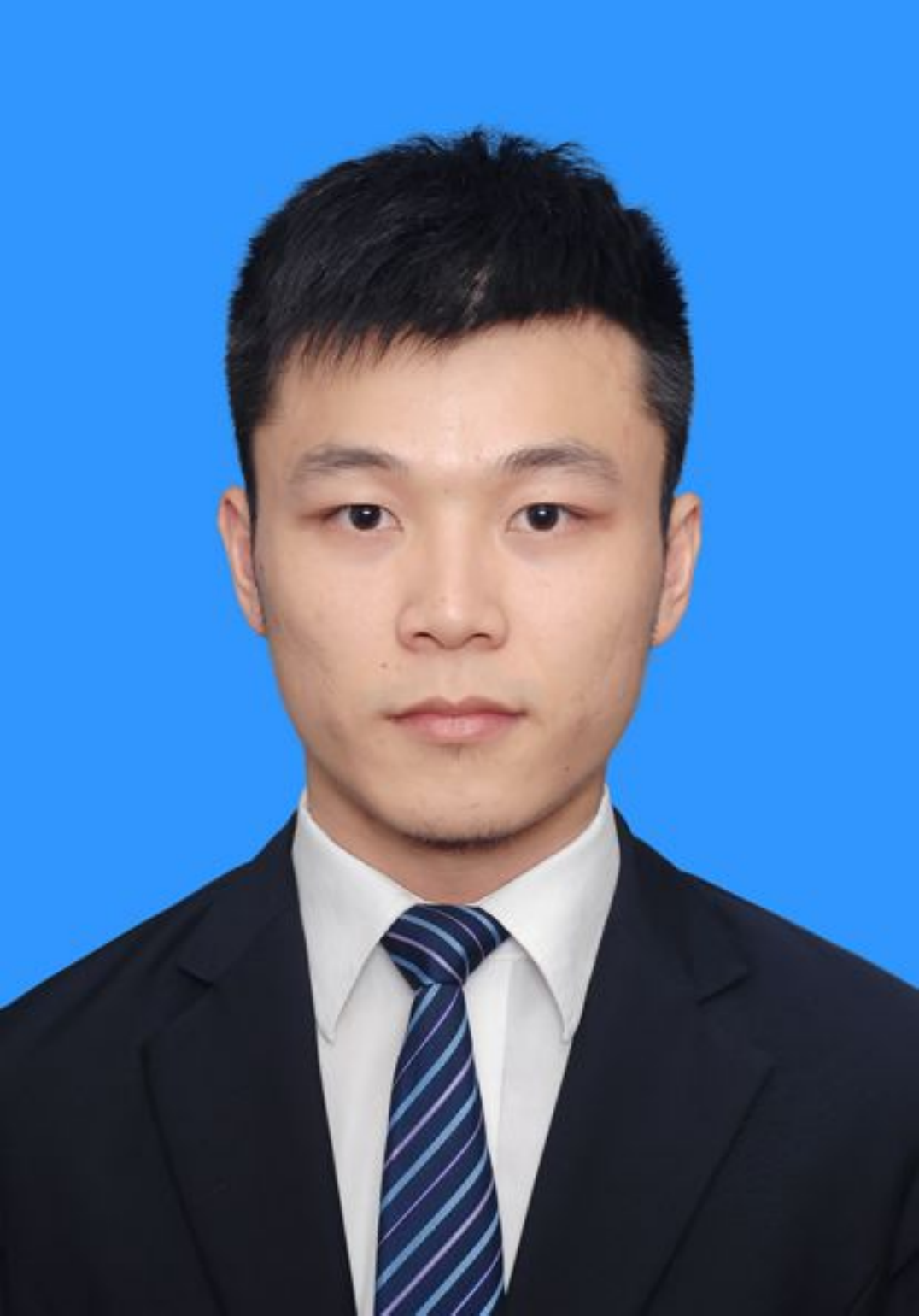}}]{Jinpeng Guo} (S'19) received the B.S. degree in electrical engineering and automation from Chongqing University, Chongqing, China, in 2014 and M.S. degree in electrical engineering from Southeast University, Nanjing, China, in 2017. He is currently pursuing the Ph.D. degree in the Department of Electrical and Computer Engineering, McGill University, Montreal, QC, Canada.  His research interests include power system monitoring, analysis and control.
\end{IEEEbiography}

\begin{IEEEbiography}[{\includegraphics[width=1in,height=1.25in,clip,keepaspectratio]{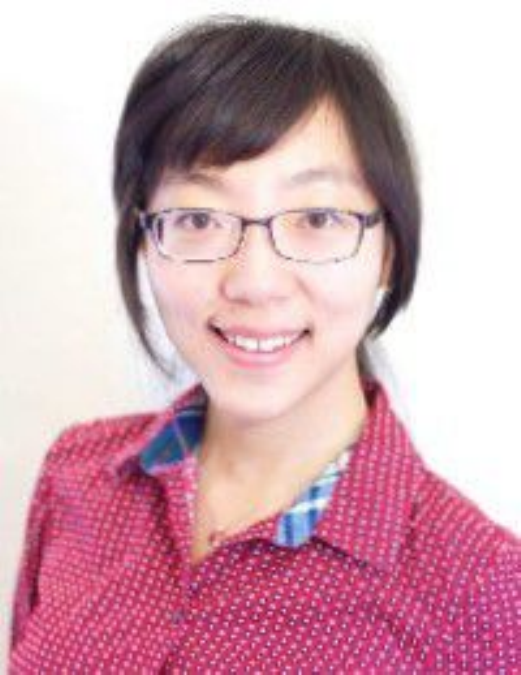}}]{Xiaozhe Wang} (S’13-M’16) is currently an Assistant Professor in the Department of Electrical and Computer Engineering at McGill University, Montreal, QC, Canada. She received the Ph.D. degree in the School of Electrical and Computer Engineering from Cornell University, Ithaca, NY, USA, in 2015, and the B.S. degree in Information Science and Electronic Engineering from Zhejiang University, Zhejiang, China, in 2010. 

Her research interests are in the general areas of power system stability and control, uncertainty quantification in power system security and stability, and wide-area measurement system (WAMS)-based detection, estimation, and control. She is serving on the editorial boards of IEEE Transactions on Power Systems, Power Engineering Letters, and IET Generation, Transmission and Distribution.
\end{IEEEbiography}

\begin{IEEEbiography}[{\includegraphics[width=1in,height=1.25in,clip,keepaspectratio]{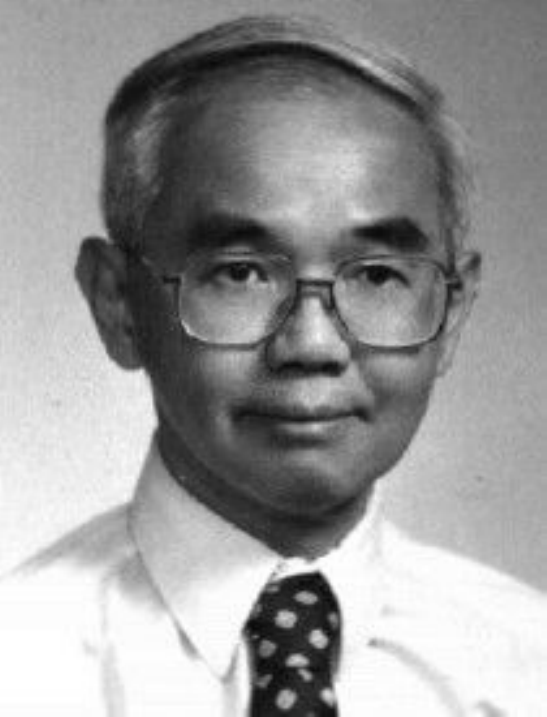}}]{Boon-Teck Ooi} (M’71-SM’85-F’02-LF’05) was born in Malaysia. He received the B. Eng. (Honors) degree in electrical engineering from the University of Adelaide, Australia, the M.S. degree in electrical engineering from the Massachusetts Institute of Technology and the Ph.D. degree in electrical engineering from McGill University, Montreal, QC, Canada. He is Emeritus Professor with the Department of Electrical and Computer Engineering, McGill University. He is IEEE Life Fellow. 

His research interests are in linear and conventional electric motors and generators (steady-state, transient, stability); power electronics (voltage-source converters, current-source converters, multi-level converters, power quality, thyristor HVDC, PWM-HVDC, multi-terminal HVDC FACTS), wind and other renewable energy sources.
\end{IEEEbiography}

\end{document}